\documentclass[twocolumn,preprintnumbers, amssymb,amsmath,aps,floatfix,prc,nofootinbib,superscriptaddress]{revtex4-1}

\usepackage{hyperref}
\usepackage[all]{hypcap}

\usepackage{graphicx}
\usepackage{amsmath,amsthm,amssymb}

\usepackage{color}

\begin{document}

\title{Longitudinal fluctuations and decorrelations of anisotropic flows at the LHC and RHIC energies}

\author{Xiang-Yu Wu}
\affiliation{Key Laboratory of Quark and Lepton Physics (MOE) and Institute of Particle Physics, Central China Normal University, Wuhan 430079, China}

\author{Long-Gang Pang}

\affiliation{Nuclear Science Division, Lawrence Berkeley National Laboratory, Berkeley, CA 94720, USA}
\affiliation{Physics Department, University of California, Berkeley, CA 94720, USA}

\author{Guang-You Qin}
\affiliation{Key Laboratory of Quark and Lepton Physics (MOE) and Institute of Particle Physics, Central China Normal University, Wuhan 430079, China}

\author{Xin-Nian Wang}
\affiliation{Key Laboratory of Quark and Lepton Physics (MOE) and Institute of Particle Physics, Central China Normal University, Wuhan 430079, China}
\affiliation{Nuclear Science Division, Lawrence Berkeley National Laboratory, Berkeley, CA 94720, USA}
\affiliation{Physics Department, University of California, Berkeley, CA 94720, USA}

\begin{abstract}

We perform a systematic study on the decorrelation of anisotropic flows along the pseudorapidity in relativistic heavy-ion collisions at the LHC and RHIC energies.
The dynamical evolution of the QGP fireball is simulated via the CLVisc (ideal) (3+1)-dimensional hydrodynamics model, with the fully fluctuating initial condition from A-Multi-Phase-Transport (AMPT) model.
Detailed analysis is performed on the longitudinal decorrelations of elliptic, triangular and quadrangular flows in terms of flow vectors, flow magnitudes and flow orientations (event planes).
It is found that pure flow magnitudes have smaller longitudinal decorrelation than pure flow orientations, and the decorrelation of flow vectors is a combined effect of both flow magnitudes and orientations.
The longitudinal decorrelation of elliptic flow has a strong and non-monotonic centrality dependence due to the initial elliptic collision geometry: smallest decorrelation in mid-central collisions.
In contrast, the decorrelations of triangular and quadrangular flows have weak centrality dependence, slightly larger decorrelations in more peripheral collisions.
Our numerical results for Pb+Pb collisions at the LHC are in good agreement with the ATLAS data, while our RHIC results predict much larger longitudinal decorrelations as compared to the LHC.
We further analyze the longitudinal structures of the AMPT initial conditions and find that the final-state longitudinal decorrelation effects are strongly correlated with the lengths of the initial string structures in the AMPT model. 
The decorrelation effects are typically larger at lower collision energies and in more peripheral collisions due to shorter lengths of the string structures in the initial states.

\end{abstract}
\maketitle

\section{Introduction}

High-energy heavy-ion collisions, such as those performed at Relativistic Heavy-Ion Collider (RHIC) and the Large Hadron Collider (LHC), provide ideal environments to create and study the strong-interaction nuclear matter under extreme temperatures and densities.
Various experimental observations and theoretical studies have demonstrated that the hot and dense nuclear matter produced in these extremely energetic nucleus-nucleus collisions is a strongly-coupled quark-gluon plasma (QGP), which behaves like a relativistic fluid with extremely low shear-viscosity-to-entropy-density-ratio \cite{Gale:2013da, Heinz:2013th, Huovinen:2013wma, Song:2013gia, Romatschke:2017ejr}.
One of the most important evidences for the formation of the strongly-coupled QGP is the strong collective flow of the QGP fireball \cite{Adams:2003zg, Aamodt:2010pa, ATLAS:2011ah, Chatrchyan:2012ta}, which is developed from the pressure gradient of the fireball and the strong interaction among the QGP constituents.

Due to event-by-event fluctuations of the initial state energy density and geometry, the collective flow of the QGP fireball is typically anisotropic in the plane transverse to the beam axis, which leads to anisotropic momentum distributions for the final state soft hardons.
To quantify the anisotropic collective flows, the flow vector $\mathbf{V}_n = v_n \exp(in\Psi_n)$ for $n$-th order flow is usually defined, where $v_n$ is the flow magnitude and $\Psi_n$ is the flow orientation (the symmetry plane of the anisotropic flow $\mathbf{V}_n$ or the event plane) \cite{Ollitrault:1992bk}.
Relativistic hydrodynamics has been very successful in describing the space-time evolution of the QGP fireball and in explaining the observed anisotropic collective flows in relativistic heavy-ion collisions at RHIC and the LHC \cite{Romatschke:2007mq, Song:2007fn, Dusling:2007gi, Huovinen:2008te, Bozek:2009dw, Chaudhuri:2009hj, Schenke:2010rr, Pang:2012he}.
In recent years, much attention has been paid to the studies of higher-order anisotropic flows \cite{Alver:2010gr, ALICE:2011ab, Qin:2010pf, Staig:2010pn, Ma:2010dv, Xu:2010du, Teaney:2010vd, Adare:2011tg, Qiu:2011iv, Bhalerao:2011yg, Floerchinger:2011qf, ATLAS:2012at, Adamczyk:2013waa, Chatrchyan:2013kba}, event-by-event flow fluctuations \cite{ALICE:2016kpq, Aad:2013xma, Gale:2012rq, Bhalerao:2014xra}, correlations of event planes \cite{Aad:2014fla, Bhalerao:2013ina, Qin:2011uw, Qiu:2012uy}, and flow correlations in terms of symmetric cummulants \cite{ALICE:2016kpq, Zhu:2016puf, Giacalone:2016afq} and nonlinear (hydrodynamics) responses \cite{Yan:2015jma, Qian:2016fpi, Acharya:2017zfg, CMS:2017ltu, Giacalone:2018wpp}, as well as anisotropies observed in small collision systems such as p+Pb collisions at the LHC \cite{Abelev:2012ola, Aad:2012gla, Chatrchyan:2013nka, Bozek:2013ska, Bzdak:2013zma, Qin:2013bha, Schenke:2014zha, Bzdak:2014dia}.
Detailed studies along these directions significantly advance our understanding on the origin of anisotropic collective flows of the fireball, and on the transport properties of the hot and dense QGP produced in high-energy nuclear collisions.

While there has been tremendous effort devoted to the study of the QGP dynamics and anisotropic flow in the transverse directions (plane), event-by-event fluctuations in the longitudinal (pseudorapidity) direction are of great importance as well \cite{Petersen:2011fp, Xiao:2012uw, Pang:2012uw, Rybczynski:2013yba, Jia:2014ysa, Jia:2014vja, Pang:2015zrq, Khachatryan:2015oea, Bozek:2015bna, Adam:2016ows, Jia:2017kdq, Bozek:2017qir, Ze-Fang:2017ppe, Pang:2018zzo}.
Many recent studies have shown that the longitudinal fluctuations can lead to sizable decorrelations of anisotropic collective flows along the pseudorapidity ($\eta$) direction; both flow magnitudes $v_n(\eta)$ and flow orientations $\Psi_n(\eta)$ are different at two different rapidities [i.e., $\mathbf{V}_n(\eta_1) \neq  \mathbf{V}_n(\eta_2)$].
These results provide a new set of important tools for constraining the models for initial states, and for studying the transport properties and evolution dynamics of the QGP.
In order to experimentally study the longitudinal decorrelation of anisotropic flows between two different rapidity bins, CMS Collaboration have first proposed a reference-rapidity-bin method \cite{Khachatryan:2015oea}.
This method measures the correlation between rapidity bins $\eta$ and $-\eta$ using the ratio of the correlation between the $\eta$ bin and the reference rapidity bin $\eta_{\rm r}$ and the correlation between the $-\eta$  bin and the reference bin $\eta_{\rm r}$.
Since a large rapidity gap can be used between $\pm \eta$ and the reference rapidity bin $\eta_{\rm r}$, this method should be able to remove a large part of short-range correlations.
Since this method uses three rapidity bins, it is also called three-rapidity-bin method.
Very recently, ATLAS Collaboration have extended the CMS three-rapidity-bin method and measured a few new longitudinal decorrelation observables based on multiple-particle correlations in two or more rapidity bins \cite{Aaboud:2017tql}.

In this work, we perform a systematic study for the longitudinal decorrelations of elliptic, triangular and quadrangular flows in relativistic heavy-ion collisions at the LHC and RHIC energies.
We utilize the CLVisc (ideal) (3+1)-dimensional hydrodynamics model \cite{Pang:2012he, Pang:2012uw, Pang:2015zrq, Pang:2018zzo} to simulate the dynamical evolution of the QGP fireball and employ A-Multi-Phase-Transport (AMPT) model \cite{Lin:2004en} to generate the initial conditions for our hydrodynamics simulations.
A detailed analysis is presented for the longitudinal decorrelation of anisotropic flows in terms of flow vectors, flow magnitudes and flow orientations.
We further study the centrality and collision-energy dependences of the longitudinal decorrelations of anisotropic flows using the slope parameters of the longitudinal decorrelation functions.
The decorrelation observables involving four rapidity bins are also studied compared to these involving three rapidity bins.
Our numerical results provide very good descriptions of the recent ATLAS data on the longitudinal decorrelation in Pb+Pb collisions at 5.02A TeV and 2.76A TeV at the LHC.
The prediction for RHIC energy is also presented and we find much larger decorrelation at RHIC compared to the LHC energies.
We further analyze the longitudinal structures of the AMPT initial conditions, and find that the longitudinal decorrelation of anisotropic flows are typically larger at lower collision energies and in more peripheral collisions; this is mainly due to the shorter lengths of the string structures (thus larger fluctuations) in the AMPT model.

The paper is organized as follows. In Sec. II, we demonstrate the setup of the event-by-event hydrodynamics model including a very brief introduction of the AMPT model for initial condition.
In Sec. III , we present in detail our numerical results for the longitudinal decorrelations of elliptic, triangular and quadrangular flows at the LHC and RHIC energies. We also compare our results with the recent ATLAS data for Pb+Pb collisions at the LHC.
Sec. IV contains our summary.

\section{Event-by-Event CLVisc (3+1)-Dimensional Hydrodynamics Simulation}

\begin{figure}[htb]
\includegraphics[width=0.5\textwidth]{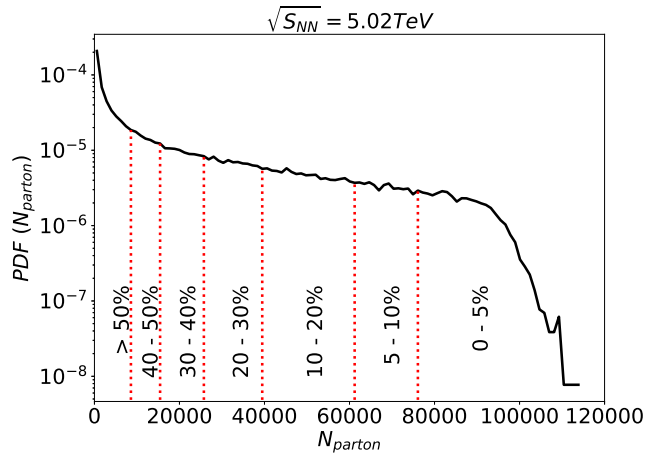}
\caption{Determination of centrality classes for Pb+Pb collisions at 5.02A~TeV according to the probability density distribution (PDF) of the initial parton multiplicity ($N_{\rm parton}$) in the AMPT model.}
\label{cent}
\end{figure}

\begin{figure*}[thb]
\includegraphics[width=1.02\textwidth]{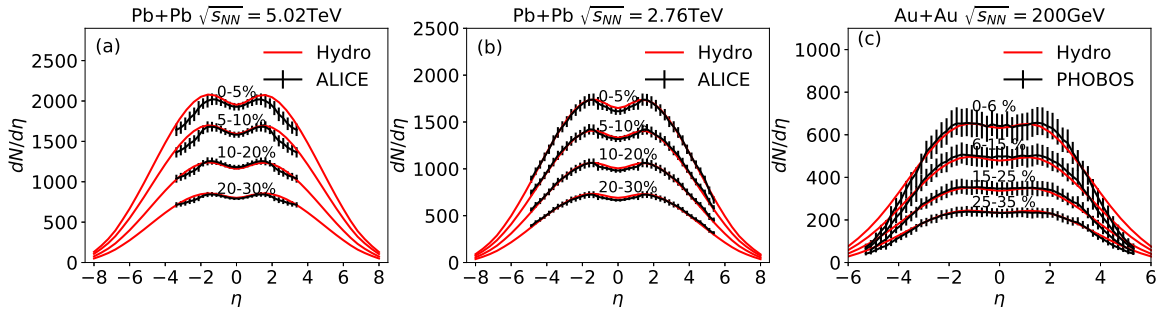}
\caption{Charged hadron multiplicity as a function of pseudorapidity $\eta$ for Pb+Pb collisions at 5.02A~TeV and 2.76A~TeV at the LHC, and for Au+Au collisions at 200A~GeV at RHIC from event-by-event (3+1)-dimensional ideal hydrodynamics simulation compared to the experimental data \cite{Adam:2016ddh, Abbas:2013bpa, Back:2002wb}.}
\label{dndeta}
\end{figure*}

In this work, we utilize the ideal version of the CLVisc (3+1)-dimensional ideal hydrodynamics model \cite{Pang:2012he, Pang:2012uw, Pang:2015zrq, Pang:2018zzo} to simulate the dynamical evolution of the QGP fireball and to study the longitudinal decorrelation of anisotropic flows in Pb+Pb collisions at $\sqrt{s_{NN}}=$2.76, 5.02 TeV and Au+Au collision at $\sqrt{s_{NN}}=$ 200GeV, respectively.
The initial conditions for hydrodynamics simulations are obtained from the AMPT model (the string-melting version) \cite{Lin:2004en}.
Using the position ($t_i,x_i,y_i,z_i $) and momentum ($E_i,P_{xi},P_{yi},P_{zi}$) information of each produced parton, we can construct the local energy-momentum tensor $T^{\mu\nu}$ at the initial proper time $\tau_0$ as follows:
\begin{eqnarray}
T^{\mu\nu} (\tau_0,x,y,\eta_s)= K \sum_i \frac{p^{\mu}_ip^{\nu}_i}{p^{\tau}_i}\frac{1}{\tau_0\sqrt{2\pi \sigma^2_{\eta_s}}}\frac{1}{2\pi \sigma^2_{r}} \times \notag \\
 \exp\left[-\frac{(x-x_i)^2+(y-y_i)^2}{2\sigma_r^2}-\frac{(\eta_s-\eta_{si})^2}{2\sigma^2_{\eta_s}}\right] \,,
\end{eqnarray}
where a normalized Guassian smearing function is applied for each parton in the Milne coordinate ($\tau,x,y,\eta_s$).
Here, $p^{\mu}$ is the four-momentum of the parton:
\begin{eqnarray}
p^{\mu} = \frac{1}{\tau_0} [m_{Ti}\text{cosh}(Y-\eta_{s}),p_{x},p_{y},m_{T}\text{cosh}(Y-\eta_{s}] \,,
\end{eqnarray}
with $Y$, $\eta_{s}$, $m_{T}$ being the rapidity, the space-time rapidity and the transverse mass of the parton, respectively.
$\sigma_r$ and $\sigma_{\eta_s}$ are the widths for the Gaussian smearing in the transverse and pseudorapidity directions, and they are taken to be $\sigma_r = 0.6$~fm and $\sigma_{\eta_s} = 0.6$ in this study.

In our initial conditions, the scale factor $K$ and the initial proper time $\tau_0$ are two key parameters.
In this study, we take the initial proper time $\tau_0=0.2$~fm for Pb+Pb collisions at 2.76A~TeV and 5.02A~GeV, and take $\tau_0=0.4$fm for Au+Au collisions at 200A~GeV.
The scale parameter $K$ is fixed by comparing the experimental data to our hydrodynamics results for charged hadron multiplicity distribution $dN/d\eta$ in most central collisions at each colliding energy.
Note that the initial flow is not included in our initial conditions.

In this study, the collision centrality classes are determined from the initial parton multiplicity distribution in the AMPT model by running one million minimum-bias events (the range of impact parameter $b \in [0,20]$~fm).
Figure \ref{cent} shows the the probability density distribution (PDF) of the initial parton multiplicity in the AMPT model and the division of the centrality classes for Pb+Pb collisions at 5.02A~TeV.
Then with initial conditions from each centrality bin, we run one thousand hydrodynamics events for our numerical analysis.
In our hydrodynamics simulations, we utilize the partial chemical equilibrium equation of state (EOS) s95p-PCE-v0 \citep{Huovinen:2009yb}.
After hydrodynamics evolution, we obtain the momentum distributions for the produced hadrons via the Cooper-Frye formula, where the freezeout temperature $T_f$ is taken to be $137$~MeV.  
Figure \ref{dndeta} shows our event-by-event hydrodynamics results on $dN_{ch}/d\eta$ in central and mid-cental collisions compared with the experimental data \cite{Adam:2016ddh, Abbas:2013bpa, Back:2002wb}.
From the comparison for central collisions, we obtain $K=1.6$ and $K=1.44$ for Pb+Pb collisions at 2.76A~TeV and 5.02A~TeV, and $K=1.45$ for Au+Au collisions.
The same $K$ values are used for non-central collisions.

\section{Numerical Results}
\label{res}

\begin{figure*}[thb]
\includegraphics[width=1.02\textwidth]{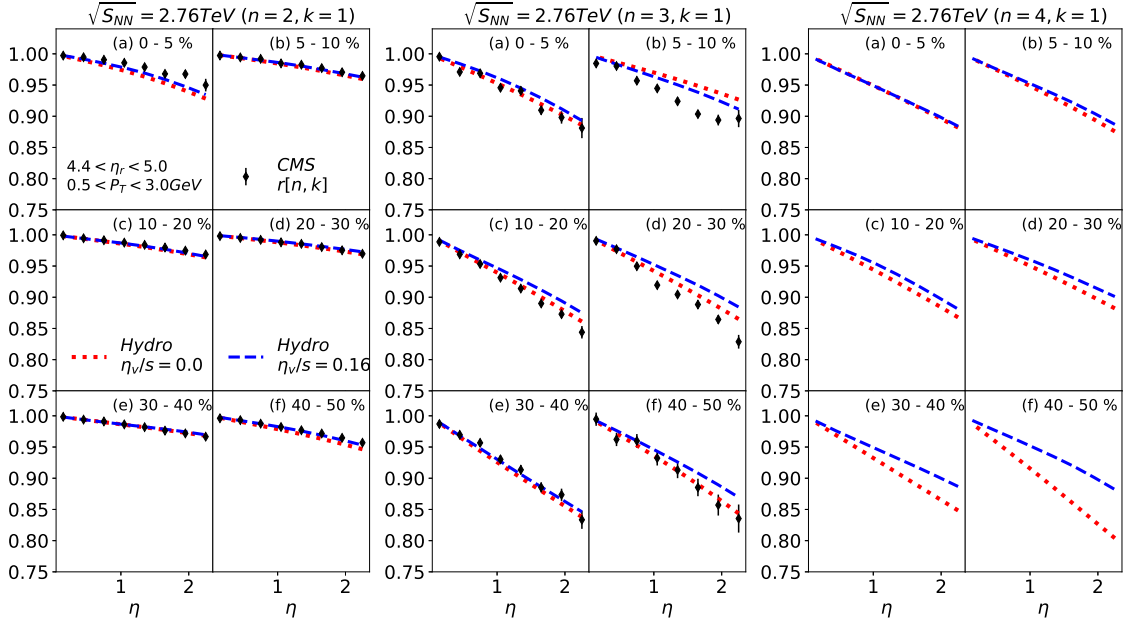}
\caption{
	The correlation functions $r[n,k](\eta)$ with $n=2, 3, 4$ and $k=1$ as a function of $\eta$ obtained from event-by-event (3+1)-D ideal hydrodynamics simulations (dotted, this work) and viscous hydrodynamics simulations (dashed, from Ref.~\cite{Pang:2018zzo}) for six different centralities in Pb+Pb collisions at 5.02A~TeV.
The CMS data for $r[2,1](\eta)$ and $r[3,1](\eta)$ are shown for comparison.
The reference rapidity window is taken to be $4.4<\eta_{\rm r}<5.0$.
	}
\label{rn_cms_2.76ATeV}
\end{figure*}

\begin{figure*}[thb]
\includegraphics[width=1.02\textwidth]{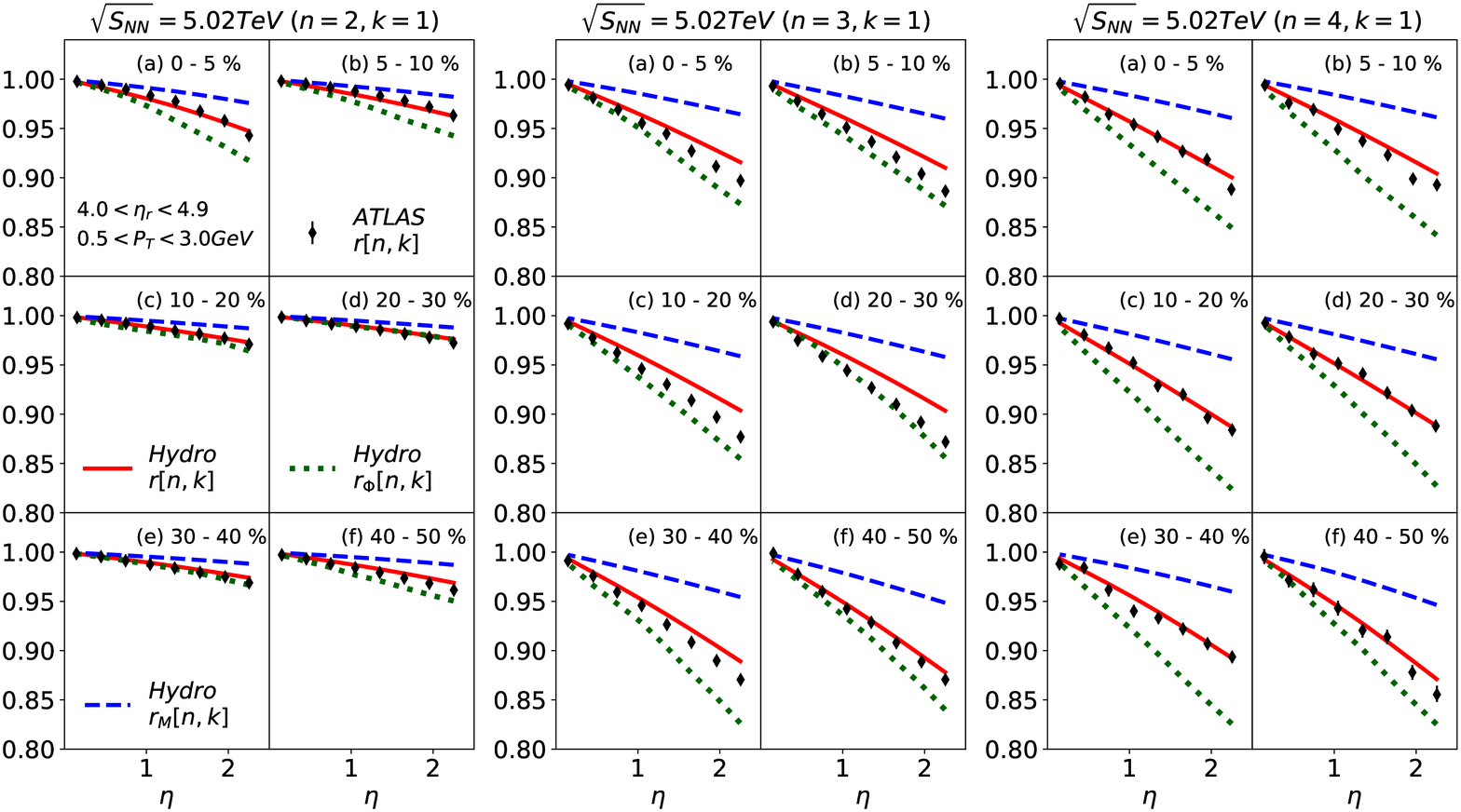}
\caption{
The correlation functions $r[n,k](\eta)$, $r_{M}[n,k](\eta)$ and $r_{\Phi}[n,k](\eta)$ with $n=2, 3, 4$ and $k=1$ as a function of $\eta$ obtained from event-by-event (3+1)-D ideal hydrodynamics simulations for six different centralities in Pb+Pb collisions at 5.02A~TeV.
The ATLAS data for $r[n,k](\eta)$ are shown for comparison.
The reference rapidity window is taken to be $4.0<\eta_{\rm r}<4.9$.
	}
\label{5020r234}
\end{figure*}

\begin{figure*}[thb]
\includegraphics[width=1.02\textwidth]{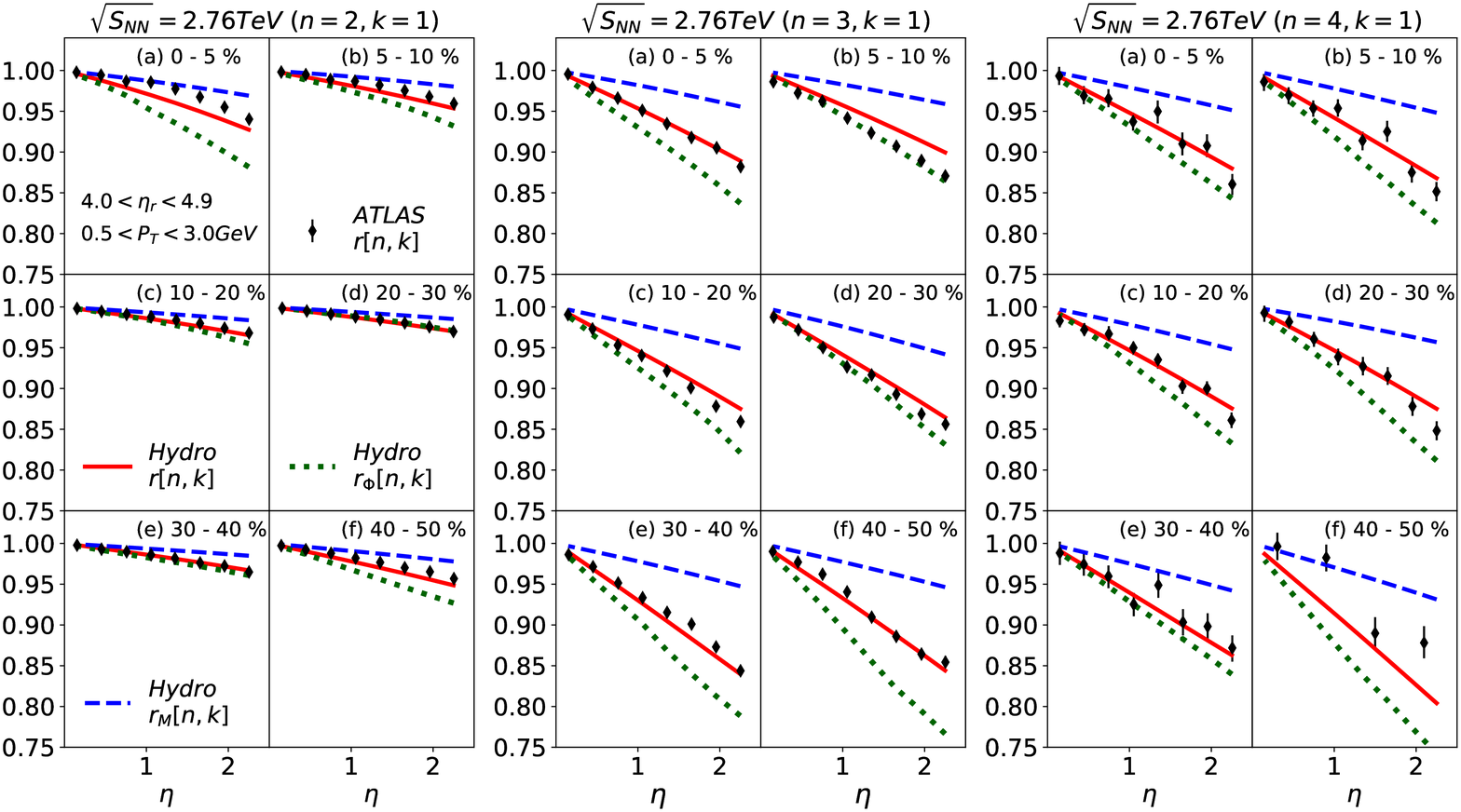}
\caption{
The correlation functions $r[n,k](\eta)$, $r_{M}[n,k](\eta)$ and $r_{\Phi}[n,k](\eta)$ with $n=2, 3, 4$ and $k=1$ as a function of $\eta$ obtained from event-by-event (3+1)-D ideal hydrodynamics simulations for six different centralities in Pb+Pb collisions at 2.76A~TeV.
The ATLAS data for $r[n,k](\eta)$ are shown for comparison.
The reference rapidity window is taken to be $4.0<\eta_{\rm r}<4.9$.
	}
\label{2760r234}
\end{figure*}

\begin{figure*}[thb]
\includegraphics[width=1.02\textwidth]{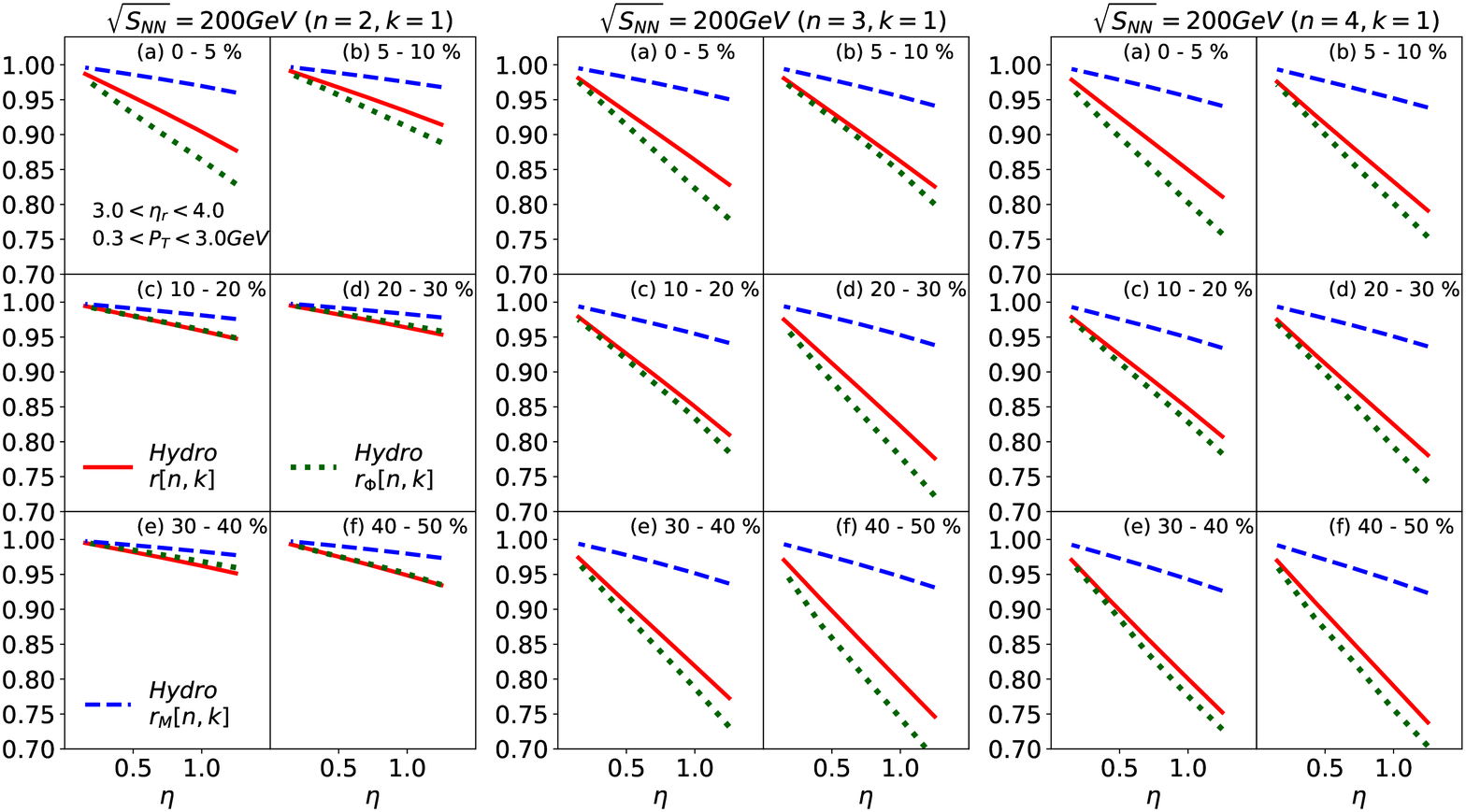}
\caption{
The correlation functions $r[n,k](\eta)$, $r_{M}[n,k](\eta)$ and $r_{\Phi}[n,k](\eta)$ with $n=2, 3, 4$ and $k=1$ as a function of $\eta$ obtained from event-by-event (3+1)-D ideal hydrodynamics simulations for six different centralities in Au+Au collisions at 200A~GeV.
The reference rapidity window is taken to be $3.0<\eta_{\rm r}<4.0$.
	}
\label{200r234}
\end{figure*}

\subsection{Longitudinal decorrelations of flow vectors, flow magnitudes and flow orientations}

In this work, we use the $\mathbf{Q}_n$ vector method to quantify the $n$-th order anisotropic collective flows in a given pseudorapidity bin:
\begin{eqnarray}
\mathbf{Q}_n(\eta) &=& q_n(\eta) e^{in\Phi_n(\eta)}=\frac{1}{N}\sum^N_{i=1} e^{in\phi_i} \,,
\end{eqnarray}
where $q_n$ and $\Phi_n$ are the magnitude and orientation of the $\mathrm{Q}_n$ vector, respectively.
Note that the $\mathbf{Q}_n$ vector constructed in the experimental measurements suffers the effects of finite multiplicity fluctuation (which are usually corrected in the experiments with the resolution factors obtained using, e.g., the sub-event method).
In our hydrodynamic simulations, we use the smooth particle spectra $\frac{dN}{d\eta p_T dp_Td\phi }$ to calculate the $\mathbf{Q}_n$ vector, i.e.,
\begin{eqnarray}
\mathbf{Q}_n(\eta) &=& \frac{\int \text{exp}(in\phi) \frac{dN}{d\eta dp_Td\phi }dp_Td\phi}{\int \frac{dN}{d\eta dp_Td\phi }dp_Td\phi} \,.
\end{eqnarray}
In this case, $\mathbf{Q}_n$ vector is the same as the flow vector $\mathbf{V}_n$.

To study the longitudinal decorrelations of anisotropic collective flows, ATLAS Collaboration \cite{Aaboud:2017tql} defines the following correlation function between the $k$-th moment of the $n$-th order flow vector in two different rapidity bins ($\eta$ and $-\eta$):
\begin{eqnarray}
r[n,k](\eta) &=& \frac{\langle \mathbf{Q}_n^k(-\eta) \mathbf{Q}_n^{*k}(\eta_{\rm r})\rangle}{\langle \mathbf{Q}_n^k(\eta)\mathbf{Q}_n^{*k}(\eta_{\rm r})\rangle} \,,
\label{origindef}
\end{eqnarray}
where the average is over many events in a given centrality class.
Note the $k=1$ case corresponds to the original three-rapidity-bin method used by the CMS Collaboration \cite{Khachatryan:2015oea}.
Here the rapidity bin $\eta$ is usually taken to be around mid-rapidity while the reference rapidity bin $\eta_{\rm r}$ is chosen at forward (large) rapidity in order to remove the short range correlations.
The above correlation function quantifies the correlation (decorrelation) between the rapidity windows $\eta$ and $-\eta$ by comparing each of them to the reference rapidity window $\eta_{\rm r}$.

In Fig.~\ref{rn_cms_2.76ATeV}, we show the numerical results for the longitudinal decorrelation functions $r[n,1](\eta)$ in Pb+Pb collisions at 2.76A TeV and compare to the CMS data (on $r[2,1]$ and $r[3,1]$).
In this figure, three sub-figures from left to right represent the longitudinal decorrelation results for elliptic flow $\mathrm{V_2}$, triangular flow $\mathrm{V_3}$ and quadrangle flow $\mathrm{V_4}$, respectively.
Within each sub-figure, six panels (plots) denote the results for six collision centralities.
Here, we compare the ideal hydrodynamics simulation results in this study and the viscous hydrodynamics simulation results from Ref. \cite{Pang:2018zzo}.
While shear viscosity may affect the overall magnitudes of the anisotropic flows $v_n(p_T, \eta)$, one can see that for most of the centralities explored here, the longitudinal decorrelation functions $r[n,k](\eta)$ for anisotropic flows are not very sensitive to shear viscosity to entropy density ratio $\eta_v/s$.

In the following, we will present a systematic study on the longitudinal decorrelations of elliptic, triangular and quadrangular flows at both the LHC and RHIC energies using (3+1)-dimensional ideal hydrodynamics.
For the LHC analysis, we follow the ATLAS setup: we limit the central rapidity windows to $\eta\in(-2.4,2.4)$, the reference rapidity bin $4.0<|\eta_{\rm r}|<4.9$, and the charged hadrons with $0.5<P_T<3.0$~GeV for Pb+Pb collisions at 2.76A~TeV and 5.02A~TeV at the LHC.
For the analysis of Au+Au collisions at 200A~GeV at RHIC, we take $\eta\in(-1.5,1.5)$, $3.0<|\eta_{\rm r}|<4.0$ and $0.3<P_T<3.0$~GeV.

We note that the above defined correlation function [Eq. (\ref{origindef})] quantifies the decorrelation between the flow vectors evaluated at two different rapidities [$\mathbf{V}_n(\eta)$ and $\mathbf{V}_n(-\eta)$].
In Ref. \cite{Bozek:2017qir}, two other similar longitudinal decorrelation functions are defined, involving only flow magnitudes $v_n$ and flow orientations $\Psi_n$, respectively.
Using the $\mathbf{Q}_n$ vector notations, they can be written as follows:
\begin{eqnarray}
r_M[n,k](\eta) &=& \frac{\langle q_n^k(-\eta) q_n^{k}(\eta_{\rm r})\rangle}{\langle q_n^k(\eta) q_n^{k}(\eta_{\rm r})\rangle} \,,
\end{eqnarray}
and
\begin{eqnarray}
r_{\Phi}[n,k](\eta) &=& \frac{\langle {\hat{Q}}_n^k(-\eta) {\hat{Q}}_n^{*k}(\eta_{\rm r})\rangle}{\langle {\hat{Q}}_n^k(\eta) {\hat{Q}}_n^{*k}(\eta_{\rm r})\rangle} \nonumber \\
	&=& \frac{\langle \cos[kn(\Phi(-\eta)-\Phi(\eta_{\rm r}))]\rangle}{\langle \cos[kn(\Phi(\eta)-\Phi(\eta_{\rm r}))]\rangle } \,,
\end{eqnarray}
where $\hat{Q}_n$ represent the unit vector $\hat{Q}_n = \mathbf{Q}_n/q_n$.
Compared to the correlation function $r[n,k](\eta)$ defined in Eq. (\ref{origindef}) which quantifies the longitudinal decorrelation for the full flow vector, $r_M$ and $r_\Phi$ describe the longitudinal decorrelation of pure flow magnitudes and pure flow orientations at two different rapidities.

In Figures \ref{5020r234}, \ref{2760r234} and \ref{200r234}, we show the numerical results for the longitudinal decorrelations of full flow vectors $r[n,k](\eta)$, flow angles $r_\Phi[n,k](\eta)$ and flow magnitudes $r_M[n,k](\eta)$, for different orders $(n=2, 3, 4)$ of anisotropic collective flows, for different collision centralities, in Pb+Pb collisions at 5.02A TeV and at 2.76A TeV, and in Au+Au collisions at 200A GeV.
For simplicity, only $k=1$ results are shown in these figures; $k>1$ results and the comparison (relation) between different $k$ values will be addressed in more details in a later subsection.
From left to right in each figure, there are three sub-figures which show the decorrelation results for elliptic flow ($n=2$), triangular flow ($n=3$) and quadrangle flow ($n=4$), respectively.
Within each sub-figure, there are six panels (plots) which show the results for six different collision centralities.
In each single panel (plot), there are three curves which represent our hydrodynamics results for the decorrelations of flow vectors $r[n,k](\eta)$, flow angles $r_\Phi[n,k](\eta)$ and flow magnitudes $r_M[n,k](\eta)$.
The experimental data from the ATLAS Collaboration on the decorrelation of flow vectors $r[n,k](\eta)$ are shown for comparison.
We can see that our hydrodynamics calculations for Pb+Pb collisions at both 2.76A TeV and 5.02A TeV at the LHC provide very nice agreements with the ATLAS data for the longitudinal decorrleation of flow vectors $r[n,1](\eta)$ for all centrality classes (these results are consistent with the previous calculations \citep{Pang:2015zrq,Pang:2018zzo}).

From Figures \ref{5020r234}, \ref{2760r234} and \ref{200r234}, we can see that the longitudinal decorrelation functions $r[n,k](\eta)$, $r_\Phi[n,k](\eta)$ and $r_M[n,k](\eta)$ for anisotropic flows $v_2$, $v_3$ and $v_4$ are almost linear in pseudorapidity $\eta$ around mid-rapidity (we will test the goodness of such linearity in the next subsection).
The approximate linearity for longitudinal decorrelations works well for all centralities in Pb+Pb collisions at 5.02A TeV and 2.76A TeV and in Au+Au collisions at 200A GeV.
Another interesting feature (also found in Ref.~\cite{Bozek:2017qir}) is that the longitudinal decorrelation of pure flow orientations (event planes) is typically larger than that of pure flow magnitudes, and the longitudinal decorrelation of full flow vectors sit between the decorrelations of flow magnitudes and flow orientations.
Such feature persists for all collision centralities at both the LHC and RHIC, which might indicate that the orientations of anisotropic flows are more sensitive to initial state longitudinal fluctuations than the flow magnitudes.

Now we focus on the centrality dependence for the longitudinal decorrelations of the anisotropic flows.
At first sight, the elliptic flow decorrelation function $r[2, 1](\eta)$ shows a non-monotonic behavior from central to peripheral collisions (it first decreases and then increases, and the decorrelation effect is weakest for 20\%-30\% central collisions at both RHIC and the LHC).
Such non-monotonic centrality dependence originates from the underlying elliptic shape of the initial collision geometry, which is the driving mechanism behind the development of elliptic flow $v_2$ in mid-central collisions, whereas the fluctuations become more important for more central and more peripheral collisions.
For triangular ($n=3$) and quadrangular ($n=4$) flows, the decorelation effects have a weak dependence on collision centrality; we observe a slight increase from central to peripheral collisions.
This is mainly due to the fact that fluctuations play more important roles in the developments of $v_3$ and $v_4$, and also fluctuations are typically larger for smaller collision systems (more peripheral collisions).
In addition, the longitudinal decorrelations of $v_3$ and $v_4$ are much larger than the decorrelation of $v_2$.

Finally we look at the collision energy dependence for the longitudinal decorrelation by comparing the results in Figures \ref{5020r234}, \ref{2760r234} and \ref{200r234}.
First, the decorrelations are much larger in Au+Au collisions at RHIC than in Pb+Pb collisions at the LHC energies.
Also, more detailed comparison between 2.76A TeV and 5.02A TeV Pb+Pb collisions at the LHC shows that the decorrelation effects are a little larger in 2.76A TeV than in 5.02A TeV.
One of the most important reasons for the above collision energy dependence is the larger initial state fluctuations in the longitudinal direction in less energetic heavy-ion collisions.
A more detailed analysis on the collision energy dependence for the longitudinal fluctuations and flow decorrelations will be presented in a later subsection.

\subsection{Slope parameters for longitudinal decorrlations}

\begin{figure*}[thb]
\includegraphics[width=1.02\textwidth]{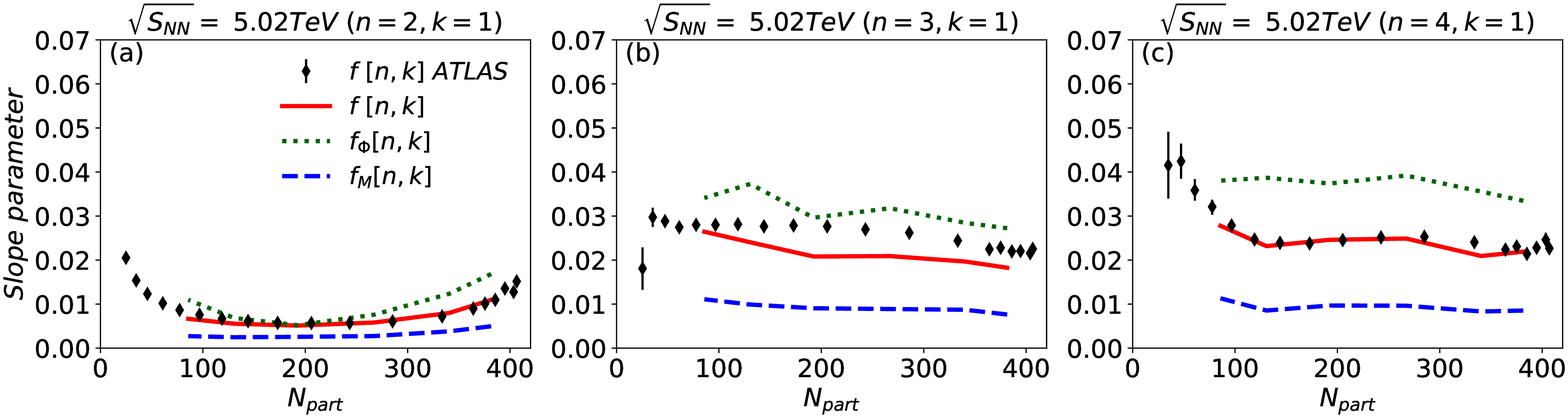}
\includegraphics[width=1.02\textwidth]{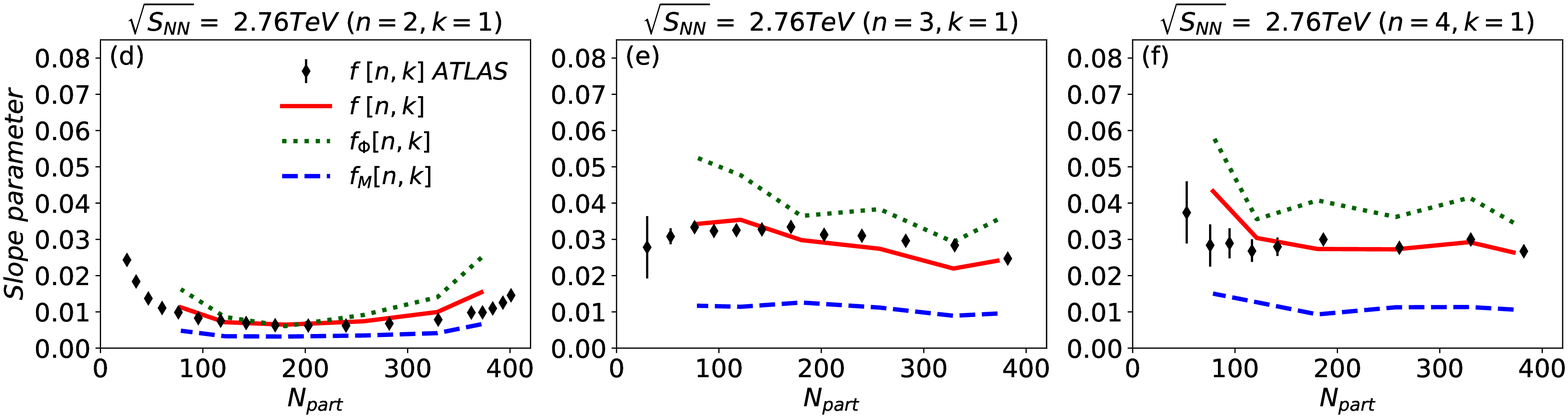}
\includegraphics[width=1.02\textwidth]{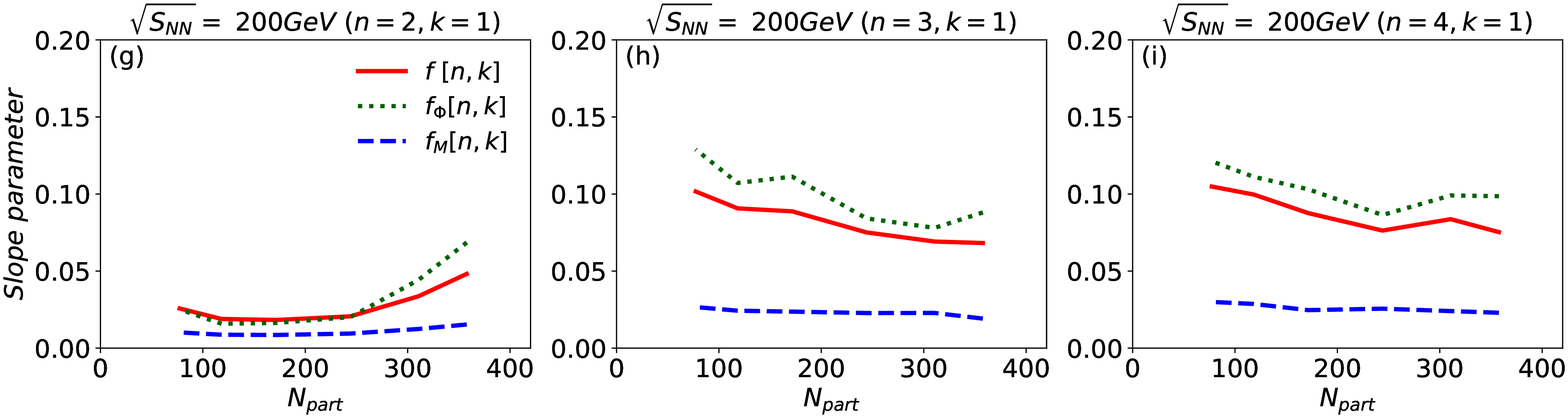}
\caption{The slope parameters $f[n,k]$, $f_M[n,k]$ and $f_\Phi[n,k]$ of the decorrelation functions $r[n,k](\eta)$, $r_M[n,k](\eta)$ and $r_\Phi[n,k](\eta)$ with $n=2, 3, 4$ and $k=1$, as a function of collision centrality ($N_{\rm part}$) for Pb+Pb collisions at 5.02A TeV and 2.76A TeV at the LHC and for Au+Au collisions at 200A GeV. The ATLAS data for 5.02A TeV Pb+Pb collisions are shown for comparison.}
\label{coffr234}
\end{figure*}

As has been shown in Figures \ref{5020r234}, \ref{2760r234} and \ref{200r234}, the longitudinal decorrelation functions for anisotropic flows, $r[n,k](\eta)$,  $r_\Phi[n,k](\eta)$ and $r_M[n,k](\eta)$ are almost linear in pseudorapidity $\eta$, especially around midrapidity. Thus it is easier to parameterize the decorrelation function $r[n,k](\eta)$ as follows:
\begin{eqnarray}
&& r[n, k](\eta) \approx 1 - 2 f[n,k] \eta
\end{eqnarray}
where $f[n,k]$ is called the slope parameter for the correlation function $r[n,k](\eta)$.
Similarly, one may define the slope parameters $f_{M}[n,k]$ and $f_{\Phi}[n,k]$ for the decorrelation functions $r_{M}[n,k](\eta)$ and $r_{\Phi}[n,k](\eta)$:
\begin{eqnarray}
&& r_{M}[n, k](\eta) \approx 1 - 2 f_{M}[n,k] \eta
\nonumber\\
&& r_{\Phi}[n, k](\eta) \approx 1 - 2 f_{\Phi}[n,k] \eta
\end{eqnarray}
In principle, these slope parameters can be directly extracted from the correlation functions $f[n,k](\eta)$, $f_{M}[n,k](\eta)$ and $f_{\Phi}[n,k](\eta)$ as a function of $\eta$ as shown in Figures \ref{5020r234}, \ref{2760r234} and \ref{200r234}.
ATLAS Collaboration measures the above slope parameters by performing the $\eta$-weighted average for the deviation of the correlation function $r(n,k)$ from the unity \cite{Aaboud:2017tql}:
\begin{eqnarray}
f[n,k] = \frac{\sum_i\left\{1 - r[n,k](\eta_i)\right\} \eta_i }{2\sum_i \eta_i^2}
\end{eqnarray}
Similarly for the slope parameters $f_{M}[n,k]$ and $f_{\Phi}[n,k]$. In this study, we use the same ATLAS method to calculate various slope parameters.

In Figure \ref{coffr234}, we show the numerical results for the slope parameters $f[n,k]$, $f_{M}[n,k]$ and $f_{\Phi}[n,k]$ of the longitudinal decorrelations of flow vectors, flow magnitudes and flow orientations as a function of centrality (the participant number $N_{\rm part}$), for different order of anisotropic flows ($v_2$, $v_3$ and $v_4$), for Pb+Pb collisions at 5.02A TeV and 2.76A TeV and Au+Au collisions at 200A GeV.
The data for the slope parameter $f[n,k]$ with $k=1$ in Pb+Pb collisions at 5.02A TeV and 2.76A TeV from ATLAS Collaboration are also shown for comparison, and we can see our calculation can describe the ATLAS data quite well.
Since the longitudinal decorrelation functions $r[n,k](\eta)$, $r_{M}[n,k](\eta)$ and $r_{\Phi}[n,k](\eta)$ are nearly linear in pseudorapidity, we can see that the slope parameters $f[n,k]$, $f_{M}[n,k]$ and $f_{\Phi}[n,k]$ can provide the same but more clear information compared to the decorrelation function. Therefore we will focus on the slope parameters in the following discussion. We will test the goodness of the linearity for the longitudinal decorrelations as a function of $\eta$ in the next subsection.

From Figure \ref{coffr234}, we can see that the longitudinal decorrelation effects are larger for pure flow orientations (event planes) and smaller for pure flow magnitudes, whereas the decorrelation of flow vector involves both flow magnitudes and orientations, and therefore sits in the middle.
Also, the decorrelation of elliptic flow shows a non-monotonic centrality dependence due to the underlying elliptic shape of the initial collision geometry, while the decorrelations for triangular and quadrangular flows just show a slight increase from central to peripheral collisions.
And the longitudinal decorrelations of $v_3$ and $v_4$ are usually much larger than the decorrelation of $v_2$.
Finally, the fluctuations and the longitudinal decorrelations of anisotropic flows in Au+Au collisions at RHIC are much larger than in Pb+Pb collisions at the LHC.

\subsection{Linearity of longitudinal decorrelations}

\begin{figure*}[thb]
\includegraphics[width=1.02\textwidth]{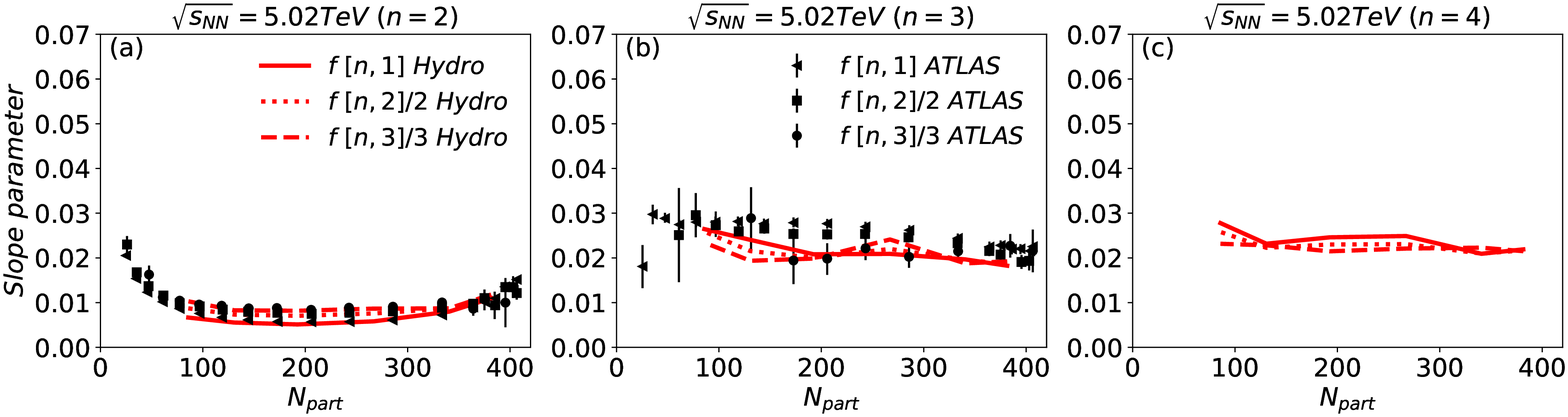}
\includegraphics[width=1.02\textwidth]{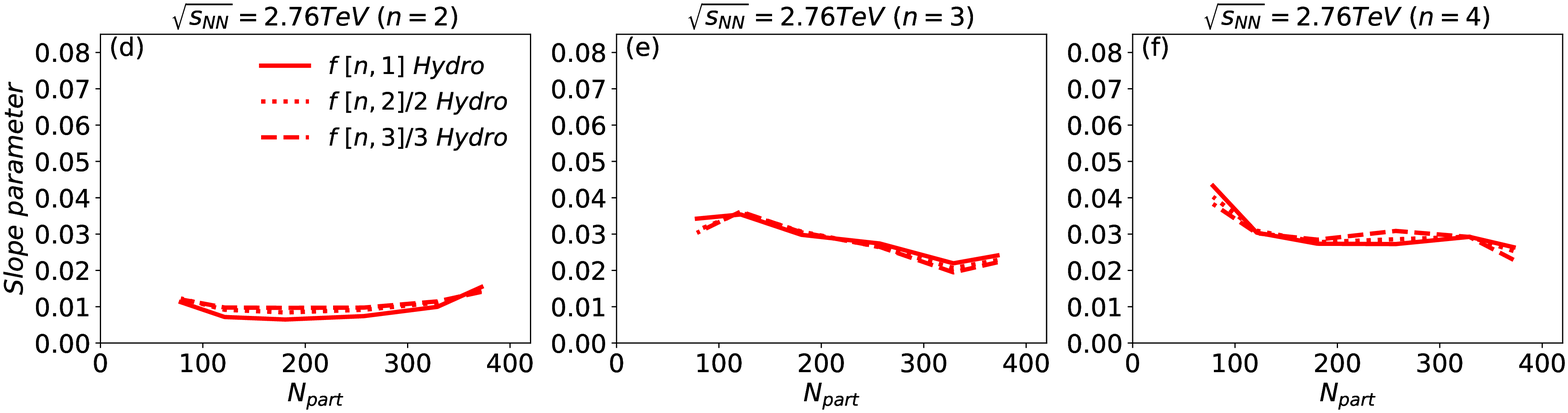}
\includegraphics[width=1.02\textwidth]{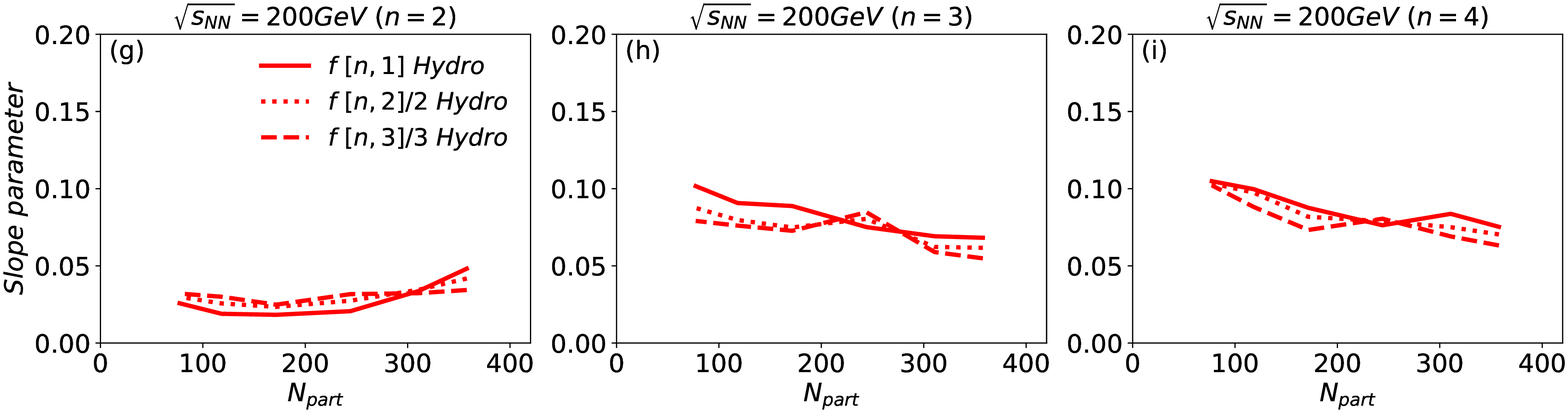}
\caption{The slope parameters $f[n,k]/k$  of the flow decorrelation functions $r[n,k](\eta)$ for different values of $n=2, 3, 4$ and different values of $k=1, 2, 3$ as a function of collision centrality ($N_{\rm part}$) for Pb+Pb collisions at 5.02A TeV and 2.76A TeV at the LHC and for Au+Au collisions at 200A GeV. The ATLAS data for 5.02A TeV Pb+Pb collisions are shown for comparison.}
\label{r234cm}
\end{figure*}

\begin{figure*}[thb]
\includegraphics[width=1.02\textwidth]{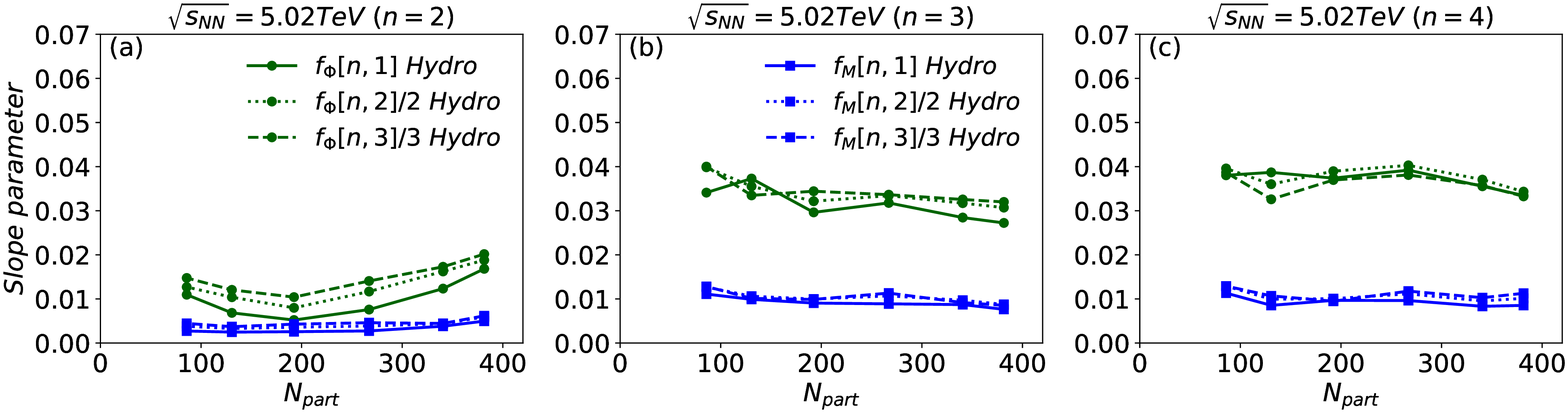}
\includegraphics[width=1.02\textwidth]{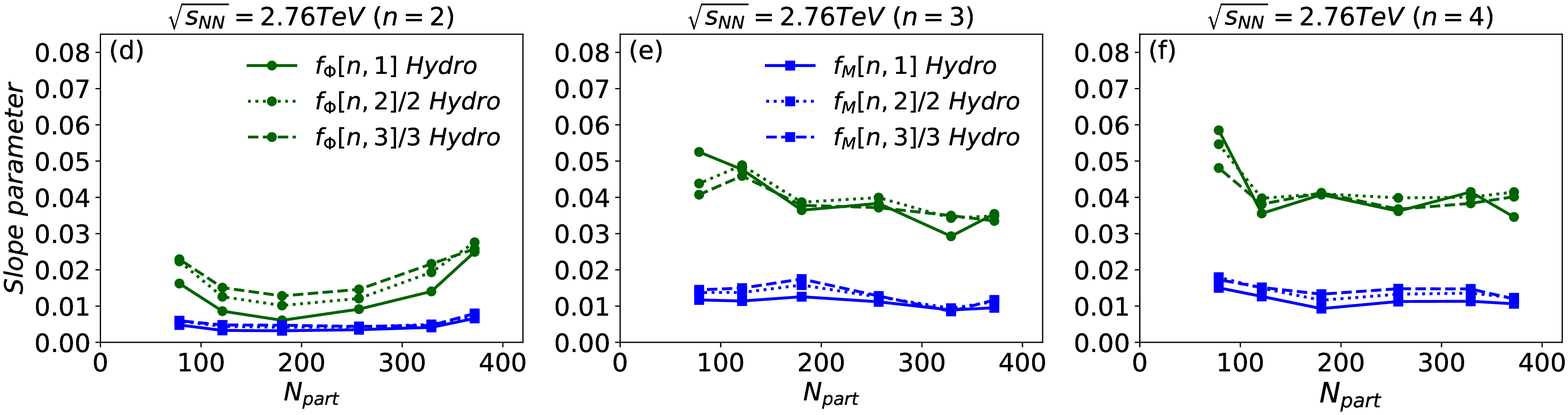}
\includegraphics[width=1.02\textwidth]{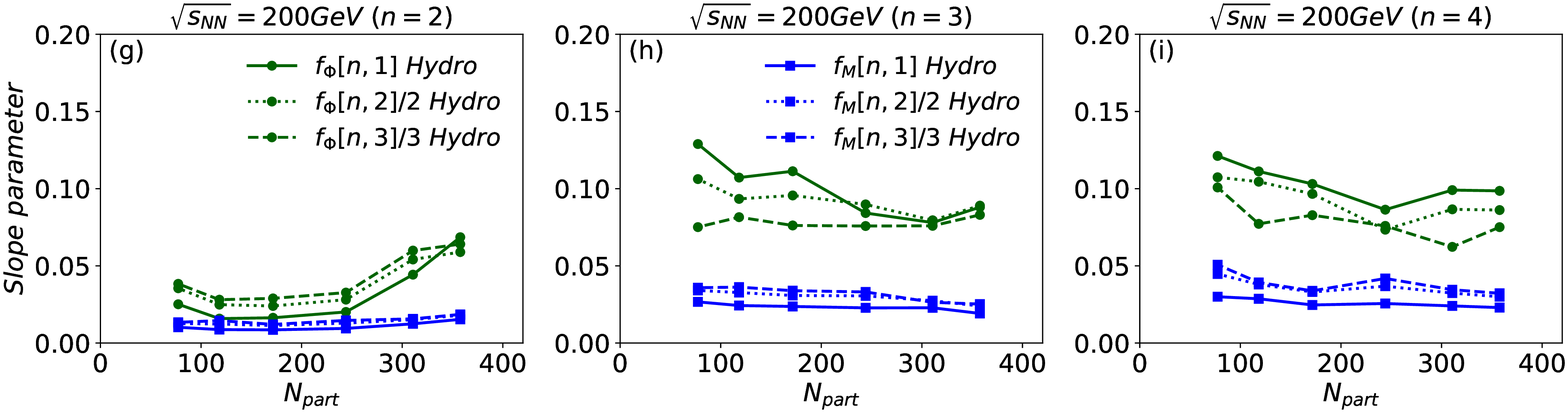}
\caption{The slope parameters $f_M[n,k]/k$ and $f_\Phi[n,k]/k$ of the flow decorrelation functions $r_M[n,k](\eta)$ and $r_\Phi[n,k](\eta)$ for different values of $n=2, 3, 4$ and different values of $k=1, 2, 3$ as a function of collision centrality ($N_{\rm part}$) for Pb+Pb collisions at 5.02A TeV and 2.76A TeV at the LHC and for Au+Au collisions at 200A GeV. }
\label{r234am}
\end{figure*}

In this subsection, we will test the validity (goodness) of the linear approximation for the longitudinal decorrelations of the anisotropic collective flows as a function of pseudorapidity $\eta$.
Following the argument of Ref. \cite{Jia:2017kdq}, we perform a perturbative expansion for the flow vector ($\mathbf{Q}$-vector) along $\eta$ direction around mid-rapidity,
\begin{eqnarray}
\mathbf{Q}_n(\eta)\approx \mathbf{Q}_n(0)(1+\alpha_n\eta)e^{i\beta_n \eta}
\end{eqnarray}
In the equation, $\mathbf{Q}_n(0)$ is the $\mathbf{Q}$-vector at midrapidity ($\eta=0$), the term $(1+\alpha_n\eta)$ characterizes the linear decorrelation along $\eta$ direction for flow magnitude, and $e^{i\beta_n\eta}$ represents the rotation (twist) of flow orientation with respect to the flow vector at mid-rapidity.
With the above approximation, the denominator of the longitudinal decorrelation function $r[n,k](\eta)$ for the flow vectors can be written as:
\begin{eqnarray}
\langle \mathbf{Q}_n^k(\eta) \mathbf{Q}_n^{*k}(\eta_{\rm r}) \rangle \approx  \langle \mathbf{Q}_n^k(0) (1+k\alpha_n \eta)e^{i  k \beta_n \eta}\mathbf{Q}_n^{*k}(\eta_{\rm r}) \rangle
\ \ \ \
\end{eqnarray}
Note the linear approximation is only valid for small $\eta$. For the flow vector $\mathbf{Q}_n(\eta_{\rm r})$ at the reference rapidity bin $\eta_{\rm r}$, one may not simply take the linear approximation since $\eta_{\rm r}$ is usually at large rapidity. Instead, we take the following parameterization:
\begin{eqnarray}
\mathbf{Q}_n^k(0)\mathbf{Q}_n^{*k}(\eta_{\rm r}) &=& A_{n,k}(\eta_{\rm r}) e^{- i \delta_{n,k}(\eta_{\rm r})} 
\nonumber\\
&=& X_{n,k}(\eta_{\rm r}) + i Y_{n,k}(\eta_{\rm r})
\end{eqnarray}
where $A_{n,k}=({X_{n,k}^2 + Y_{n,k}^2})^{1/2}$ and $e^{-i \delta_{n,k}}$ are the magnitude and the phase of the above product, respectively.
Then the denominator of the longitudinal decorrelation function $r[n,k](\eta)$ for the flow vectors can be written as follows:
\begin{eqnarray}
\langle \mathbf{Q}_n^k(\eta) \mathbf{Q}_n^{*k}(\eta_{\rm r}) \rangle
&& \approx \langle X_{n,k}(\eta_{\rm r}) \rangle 
\\
&& \times
\left[1+ k \left(\frac{\langle \alpha_n X_{n,k}\rangle}{\langle X_{n,k} \rangle} +\frac{\langle \beta_n Y_{n,k} \rangle}{\langle X_{n,k}\rangle}\right) \eta \right] \nonumber
\end{eqnarray}
With the above approximation, the decorrelation function $r[n,k](\eta)$ can be approximated as:
\begin{eqnarray}
r[n,k](\eta) \approx 1 - 2k\left(\frac{\langle \alpha_n X_{n,k} \rangle}{\langle X_{n,k} \rangle} +\frac{\langle \beta_n Y_{n,k}\rangle}{\langle X_{n,k} \rangle}\right)  \eta
\end{eqnarray}
We can see the decorrelation function $r[n,k](\eta)$ is approximately linear in pseudorapidity $\eta$, i.e.,
\begin{eqnarray}
r[n,k](\eta) \approx 1 - 2 f[n,k] \eta
\end{eqnarray}
The slope parameter $f[n,k]$ can be identified as:
\begin{eqnarray}
f[n,k] = k\left(\frac{\langle \alpha_n X_{n,k} \rangle}{\langle X_{n,k} \rangle} +\frac{\langle \beta_n Y_{n,k}\rangle}{\langle X_{n,k} \rangle}\right)
\label{slope_parameter}
\end{eqnarray}
We can do similar treatment for the longitudinal decorrelation functions $r_{M}[n,k](\eta)$ and $r_{\Phi}[n,k](\eta)$ as well. 
First, the denominator of the decorrelation function $r_{M}[n,k](\eta)$ can be written as:
\begin{eqnarray}
\langle q_n^k(\eta)q_n^{*k}(\eta_{\rm r}) \rangle 
&\approx& \langle A_{n,k} \rangle \left(1+ k \frac{\langle \alpha_n A_{n,k} \rangle}{\langle A_{n,k} \rangle} \eta\right)
\end{eqnarray}
Then the decorrelation function $f_{M}[n,k]$ becomes:
\begin{eqnarray}
	r_M[n,k](\eta) \approx 1 - 2 f_M[n,k] \eta
\end{eqnarray}
The slope parameter $f_M[n,k]$ is:
\begin{eqnarray}
f_M[n,k] = k \frac{\langle \alpha_n A_{n,k} \rangle}{\langle A_{n,k} \rangle}
\end{eqnarray}
We can see that the decorrelation function $r_M[n,k](\eta)$ only involves the slope parameter $f_M[n,k]$.
As for the decorrelation function $r_{\Phi}[n,k](\eta)$, its denominator can be written as:
\begin{eqnarray}
\langle \hat{Q}_n^k(\eta)\hat{Q}_n^{*k}(\eta_{\rm r}) \rangle
&\approx& \langle \cos(\delta_{n,k}) \rangle \left[1 + k \frac{\langle \beta_n \sin(\delta_{n,k}) \rangle}{\langle \cos(\delta_{n,k} \rangle)} \eta\right] \ \ \ \ 
\end{eqnarray}
Then the decorrelation function $r_{\Phi}[n,k]$  becomes:
\begin{eqnarray}
	r_\Phi[n,k] \approx 1 - 2 f_\Phi[n,k] \eta
\end{eqnarray}
The slope parameter $f_\Phi[n,k]$ is:
\begin{eqnarray}
f_{\Phi}[n,k] = k \frac{\langle \beta_n \sin(\delta_{n,k}) \rangle}{\langle \cos(\delta_{n,k} \rangle)}
\end{eqnarray}
We can see that the slope parameter $f_{\Phi}[n,k]$ only involves the decorrelations of flow orientations.

The above analysis tells that if the linear approximation works well for the longitudinal decorelation functions as a function of pseudorapidity $\eta$, we will have the following simple approximate relations between the slope parameters with different $k$ values:
\begin{eqnarray}
&& f[n,k] \approx k f[n,1] \nonumber\\
&& f_{M}[n,k] \approx k f_{M}[n,1] \nonumber\\
&& f_{\Phi}[n,k] \approx k f_{\Phi}[n,1]
\end{eqnarray}
These relations can be used to test the goodness of the linearity of the longitudinal decorrelation functions.

In Figure \ref{r234cm}, we show the (scaled) slope parameters $f[n,k]/k$ for the decorrelation functions $r[n,k](\eta)$, with $n=2, 3, 4$ and $k=1, 2, 3$, as a function of collision centrality ($N_{\rm part}$) for Pb+Pb collisions at 5.02A TeV and 2.76A TeV and for Au+Au collisions at 200A GeV. First, our hydrodynamics calculation provide a nice description of the ATLAS data for the slope parameters $f[n,k]/k$ in 5.02A TeV Pb+Pb collisions.
Second, three sets of curves with $k=1,2,3$ in Au+Au collisions at RHIC and Pb+Pb collisions at the LHC roughly agree with each other, indicating that the longitudinal decorrelation function $r[n,k](\eta)$ can be well approximated by a linear function of the pseudorapidity $\eta$.

In Figure \ref{r234am}, we show the slope parameters $f_M[n,k]/k$ and $f_\Phi[n,k]/k$ for the decorrelation functions $r_M[n,k](\eta)$ and $r_\Phi[n,k](\eta)$, with $n=2, 3, 4$ and $k=1, 2, 3$, as a function of collision centrality ($N_{\rm part}$) for Pb+Pb collisions at 5.02A TeV and 2.76A TeV at the LHC and for Au+Au collisions at 200A GeV.
One can see that three sets of curves (solid, dotted and dashed for $k=1,2,3$, respectively) agree quite well with each other, for both $f_M[n,k]$ and $f_\Phi[n,k]$, for elliptic, triangular and quadragular flows, at both RHIC and the LHC energies. 
This is very similar to Figure \ref{r234cm}, which means that the linear function in $\eta$ is also good approximation for $r_M[n,k](\eta)$ and $r_\Phi[n,k](\eta)$.
We should point out that there is also some breaking of the lineariality in $\eta$ for the longitudinal decorrelation functions $r[n,k](\eta)$, $r_M[n,k](\eta)$ and $r_\Phi[n,k](\eta)$ as shown by Figure \ref{r234cm} and Figure \ref{r234am}.

\subsection{Four-rapdity-bin decorrelation observables}

\begin{figure*}[thb]
\includegraphics[width=1.02\textwidth]{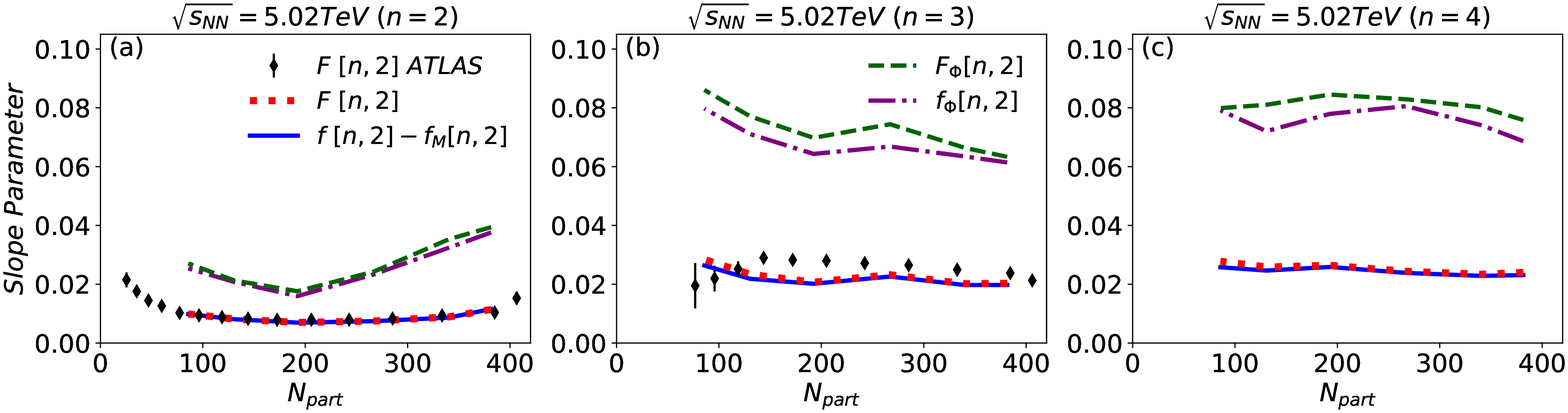}
\includegraphics[width=1.02\textwidth]{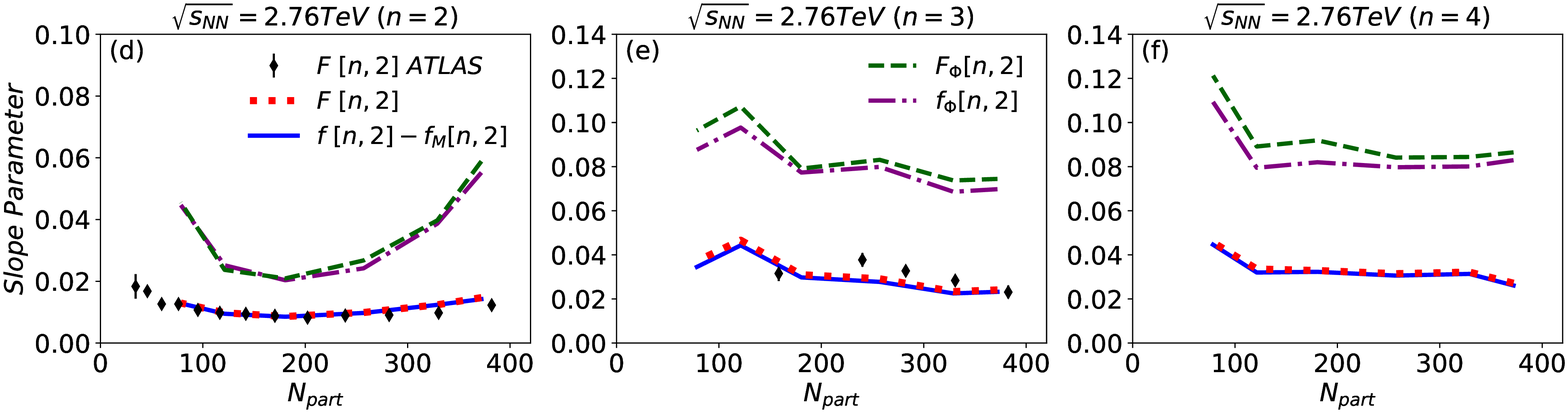}
\includegraphics[width=1.02\textwidth]{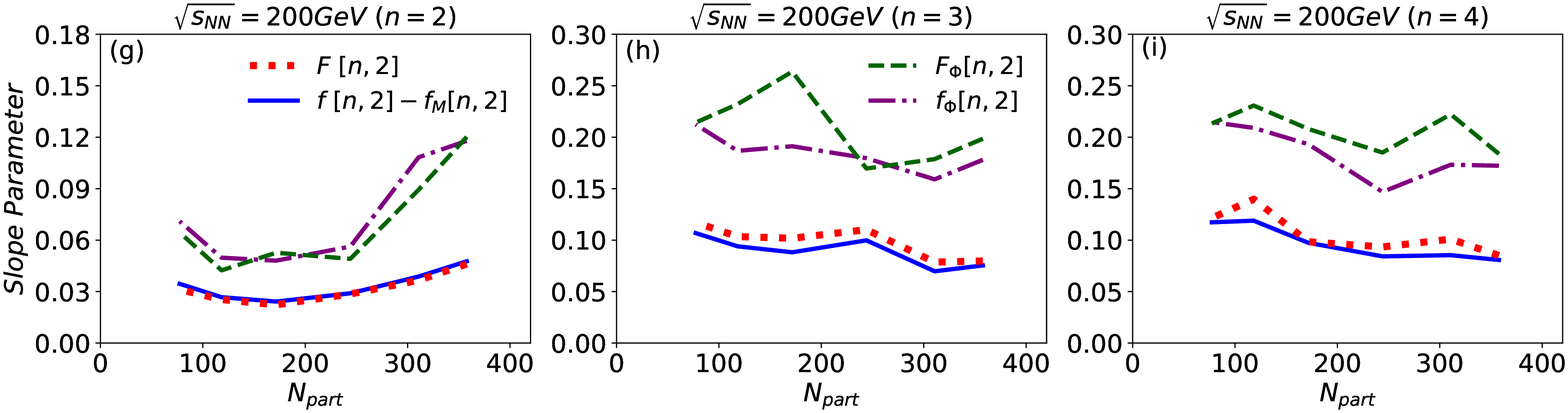}
\caption{Comparison of the slope parameters $F[n,2]$ and $F_{\Phi}[n,2]$ extracted from $R[n,2](\eta)$ with the slope parameters $f[n,2]-f_{M}[n,2]$ and $f_{\Phi}[n,2]$ extracted from $r[n,2](\eta)$, $r_M[n,2](\eta)$ and $r_\Phi[n,2](\eta)$, for different values of $n=2, 3, 4$, as a function of centrality ($N_{\rm part}$) for Pb+Pb collisions at 5.02A TeV and 2.76A TeV at the LHC and for Au+Au collisions at 200A GeV. }
\label{relationr4}
\end{figure*}

\begin{figure*}[thb]
\includegraphics[width=1.02\textwidth]{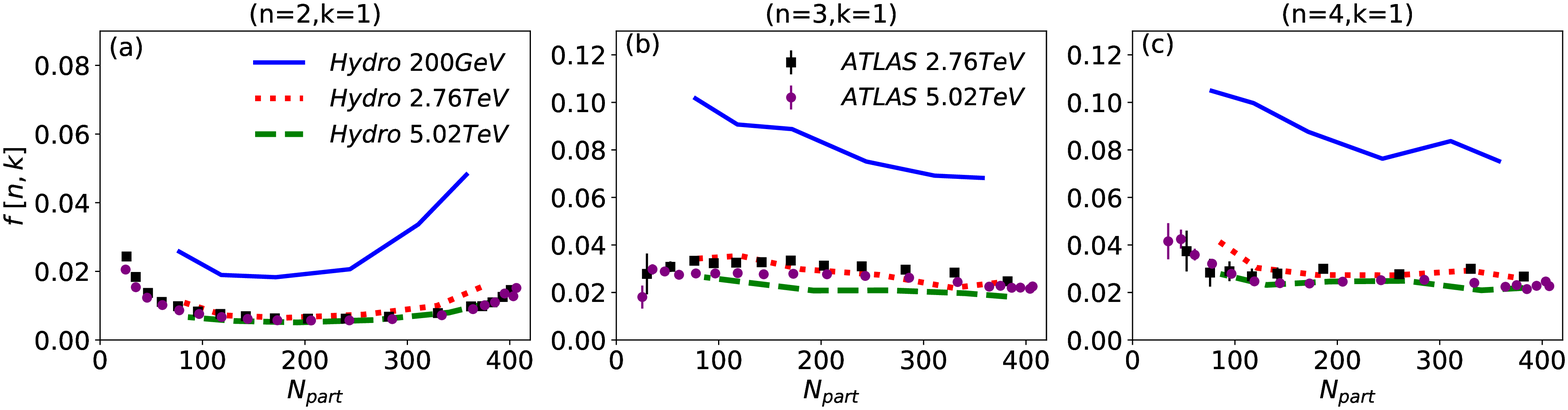}
\includegraphics[width=1.02\textwidth]{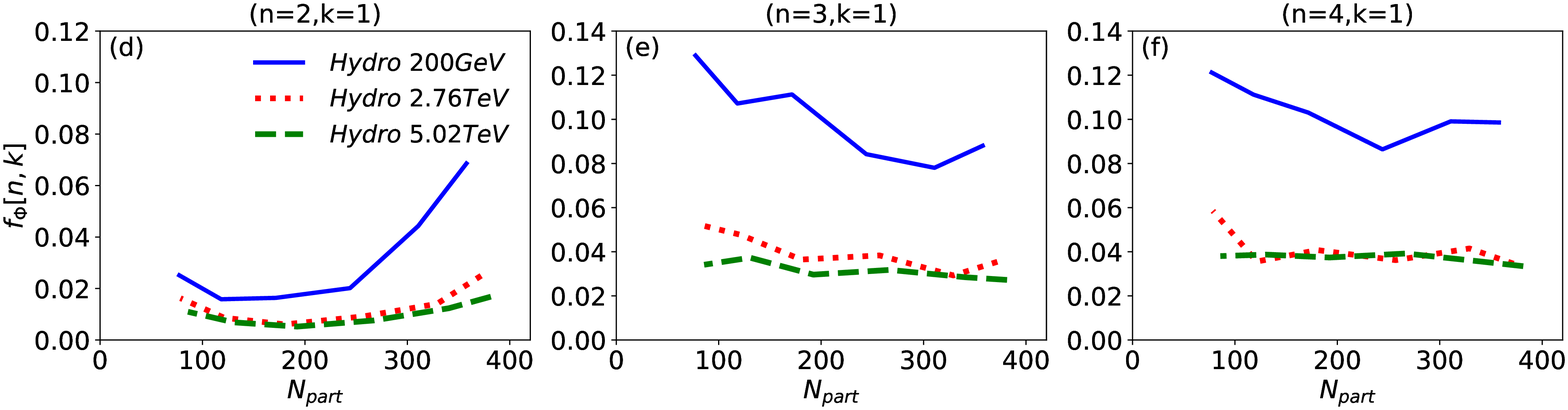}
\includegraphics[width=1.02\textwidth]{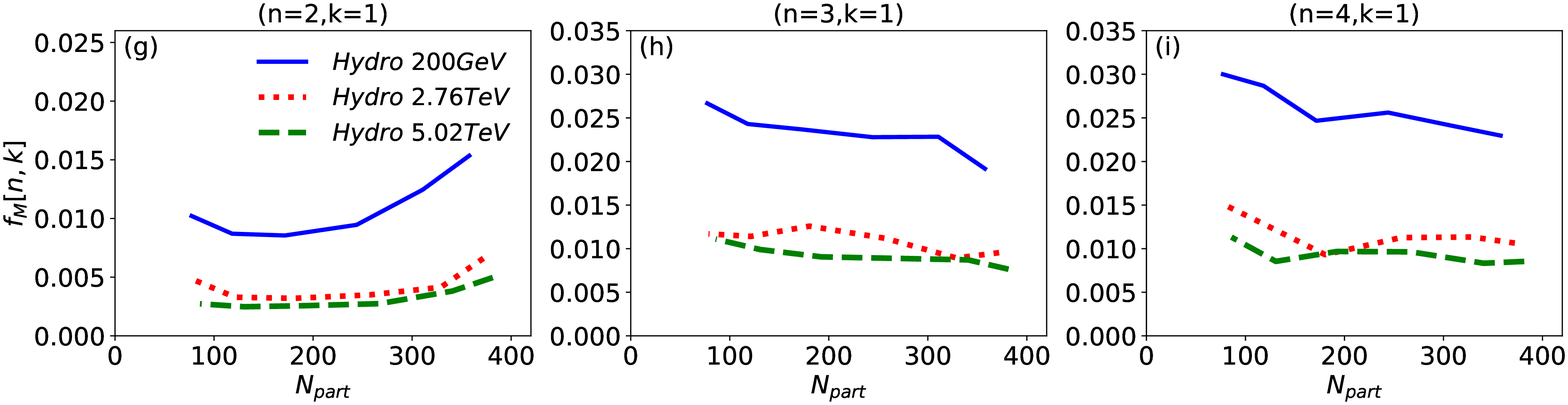}
\caption{The slope parameters $f[n,k]$ (with $n=2,3,4$ and $k=1$) as a function of centrality ($N_{\rm part}$) for Pb+Pb collisions at 5.02A TeV and 2.76A TeV at the LHC and for Au+Au collisions at 200A GeV. The ATLAS data on $f[n,k]$ are compared.}
\label{r234enr}
\end{figure*}

As we have seen from Eq. (\ref{slope_parameter}), the slope parameter $f[n,k]$ of the correlation function $r[n,k](\eta)$ contains two separate contributions: 
\begin{eqnarray}
f[n,k]  = f_{asy}[n,k] + f_{twi}[n,k]
\end{eqnarray}
with
\begin{eqnarray}
f_{asy}[n,k] = k \frac{\langle \alpha_n X_{n,k} \rangle}{\langle X_{n,k} \rangle}\,,
f_{twi}[n,k] = k \frac{\langle \beta_n Y_{n,k} \rangle}{\langle X_{n,k} \rangle}
\end{eqnarray}
These two contributions have been interpreted as the contribution from the forward-backward asymmetry ($f_{asy}[n,k]$) and the contribution from the event-plane twist ($f_{twi}[n,k]$) \cite{Jia:2017kdq}. 
It has been argued in Ref. \cite{Bozek:2017qir} that the slope parameter $f_{asy}[n,k]$ can be approximately measured via the slope parameter $f_M[n,k]$:
\begin{eqnarray}
f_{asy}[n,k] = k \frac{\langle \alpha_n X_{n,k} \rangle}{\langle X_{n,k} \rangle} 
	\approx k \frac{\langle \alpha_n A_{n,k} \rangle}{\langle A_{n,k} \rangle} = 
	f_M[n,k]
\end{eqnarray}
Here we focus on the $k=2$ case and study the following correlation functions: 
\begin{eqnarray}
&& r[n,2](\eta) = 1 - 2 f[n,2]\eta = 1 - 2 (f_{asy}[n,2] + f_{twi}[n,2]) \eta \nonumber\\
&& r_M[n,2](\eta) = 1 - 2 f_M[n,2] \eta
\end{eqnarray}
The relation $f_{asy}[n,k] \approx f_M[n,k]$ means that by measuring the correlation functions $r[n,2](\eta)$ and $r_M[n,2](\eta)$, 
we can extract the slope parameter $f_{twi}[n,2]$:
\begin{eqnarray}
f_{twi}[n,2] \approx f[n,2]-f_M[n,2]
\end{eqnarray}
In fact, the slope parameters $f_{twi}[n,2]$ can be measured from a new observable $R[n,2](\eta)$ proposed by the ATLAS Collaboration \cite{Aaboud:2017tql,Jia:2014vja}.
\begin{eqnarray}
	R[n,2](\eta) &=& \frac{\langle \mathbf{Q}_n(-\eta_{r})\mathbf{Q}_n(-\eta)\mathbf{Q}^{*}_n(\eta)\mathbf{Q}^{*}_n(\eta_{r})\rangle}{\langle \mathbf{Q}_n(-\eta_{r})\mathbf{Q}^*_n(-\eta)\mathbf{Q}_n(\eta) \mathbf{Q}^{*}_n(\eta_{r}) \rangle}
\end{eqnarray}
One can see that this new observable involves four rapidity bins.
Now we may perform similar linar approximation analysis. 
We first look at the denominator of the correlation function $R[n,2](\eta)$:
\begin{eqnarray}
\langle \mathbf{Q}_n(-\eta_{\rm r}) \mathbf{Q}_n^*(-\eta)  \mathbf{Q}_n(\eta)\mathbf{Q}_n^*(\eta_{\rm r})\rangle
&\approx& \langle X_{n,2} \rangle 
\\
&\times&
\left(1 + 2\frac{\langle \beta_n Y_{n,2} \rangle}{\langle X_{n,2} \rangle} \eta\right)
\nonumber
\end{eqnarray}
Then the correlation function $R[n,2](\eta)$ becomes:
\begin{eqnarray}
	R[n,2](\eta) = 1 - 2 F[n,2] \eta
\end{eqnarray}
The slope parameter $F[n,2]$ is:
\begin{eqnarray}
F[n,2] = 2\frac{\langle \beta_n Y_{n,2} \rangle}{\langle X_{n,2} \rangle} = f_{twi}[n,2]
\end{eqnarray}
It is clear that if the decorrelation functions $r[n,2](\eta)$ and $R[n,2](\eta)$ are linear functions of $\eta$, we have the following relation between the slope parameters:
\begin{eqnarray}
	F[n,2] 
	&\approx& f[n,2] - f_M[n,2]
\end{eqnarray}
The above relation can be tested by measuring the longitudinal decorrelation functions $R[n,2](\eta)$, $r[n,2](\eta)$ and $r_M[n,2](\eta)$, respectively.  
Similarly, we may define the following four-rapidity-bin decorrelation function $R_{\Phi}[n,2]$:
\begin{eqnarray}
	R_\Phi[n,2](\eta) &=& \frac{\langle \hat{Q}_n(-\eta_{r})\hat{Q}_n(-\eta)\hat{Q}^{*}_n(\eta)\hat{Q}^{*}_n(\eta_{r})\rangle}{\langle \hat{Q}_n(-\eta_{r})\hat{Q}^*_n(-\eta)\hat{Q}_n(\eta) \hat{Q}^{*}_n(\eta_{r}) \rangle} \ \ \ \ 
\end{eqnarray}
This observable involves only the decorrelations of flow orientations.
Performing the linear approximation analysis, its denominator becomes:
\begin{eqnarray}
&& \langle \hat Q_n(-\eta_{r}) \hat Q_n^*(-\eta) \hat Q_n(\eta)\hat Q^{*}_n(\eta_{r})\rangle 
\nonumber\\
&& \approx \langle \cos(\delta_{n,2}) \rangle \left[1 + 2 \frac{\langle \beta_n \sin(\delta_{n,2}) \rangle}{\langle \cos(\delta_{n,2} \rangle)} \eta\right]
\end{eqnarray}
Then the correlation function $R_{\Phi}[n,2](\eta)$ becomes: 
\begin{eqnarray}
	R_\Phi[n,2](\eta) \approx  1 - 2 F_{\Phi}[n,2] \eta
\end{eqnarray}
The slope parameter $F_{\Phi}[n,2]$ is:
\begin{eqnarray}
F_{\Phi}[n,2] = 2 \frac{\langle \beta_n \sin(\delta_{n,2}) \rangle}{\langle \cos(\delta_{n,2} \rangle)}  = f_\Phi[n,2]
\end{eqnarray}
We can see that both $r_{\Phi}[n,2](\eta)$ and $R_{\Phi}[n,2](\eta)$ measure the (same) decorrelation of pure flow orientations.

Figure \ref{relationr4} shows our hydrodynamics results for different slope parameters extracted from three-rapidity-bin and four-rapidity-bin correlation functions. In particular, we compare $F[n,2]$ with $f[n,2]-f_{M}[n,2]$ and $F_{\Phi}[n,2]$ with $f_{\Phi}[n,2]$, for $n=2, 3, 4$, as a function of centrality ($N_{\rm part}$) for Pb+Pb collisions at 5.02A TeV and 2.76A TeV and for Au+Au collisions at 200A GeV.
We can see that the relation $F[n,2] = f_{twi}[n,2] = f[n,2] - f_{asy}[n,2] \approx f[n,2] - f_M[n,2]$ holds pretty well for Pb+Pb collision at two LHC energies, and for Au+Au collisions at RHIC there is a very small violation of this relation. 
This means that the decorrelation functions $R[n,2](\eta)$, $r[n,2](\eta)$ and $r_M[n,2](\eta)$ at the LHC energies are well described by the linear functions in pseudorapidity $\eta$, and such linearity is slightly violated at RHIC.
One the other hand, there is sizable violation for the relation $F_{\Phi}[n,2] \approx f_{\Phi}[n,2]$, and the violation is typically larger at RHIC than at the LHC energies.
This is because the correlation functions $R_\Phi[n,2]$ and $r_\Phi[n,2]$ describe the decorrelations of pure flow orientations whose effects are typically much larger than the decorrelations of flow magnitudes.

\subsection{Collision energy and centrality dependences}

\begin{figure*}[thb]
\includegraphics[width=1.02\textwidth]{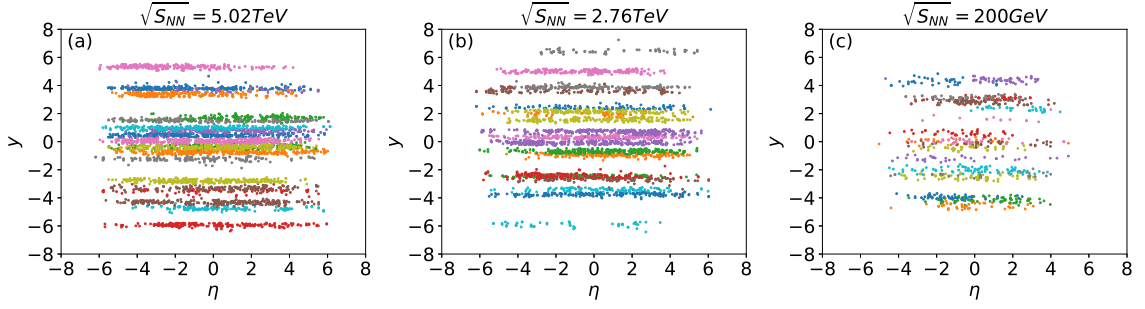}
\caption{Strings formed by the initial patrons for some typical events in Pb+Pb collisions at 5.02A TeV and 2.76A TeV and for Au+Au collisions at 200A GeV.}
\label{length}
\end{figure*}

\begin{figure*}[thb]
\includegraphics[width=1.02\textwidth]{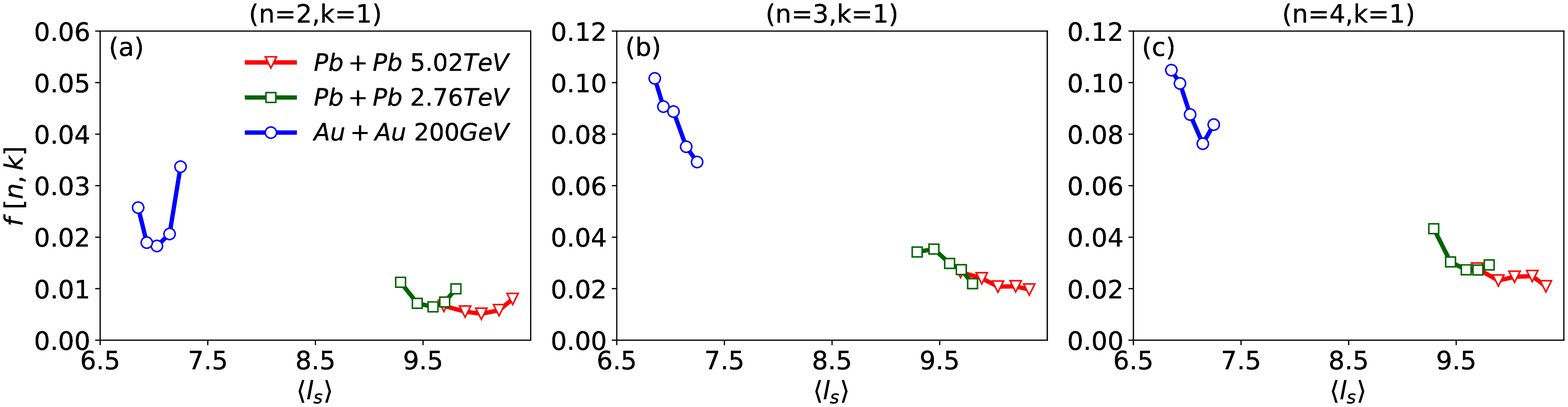}
\includegraphics[width=1.02\textwidth]{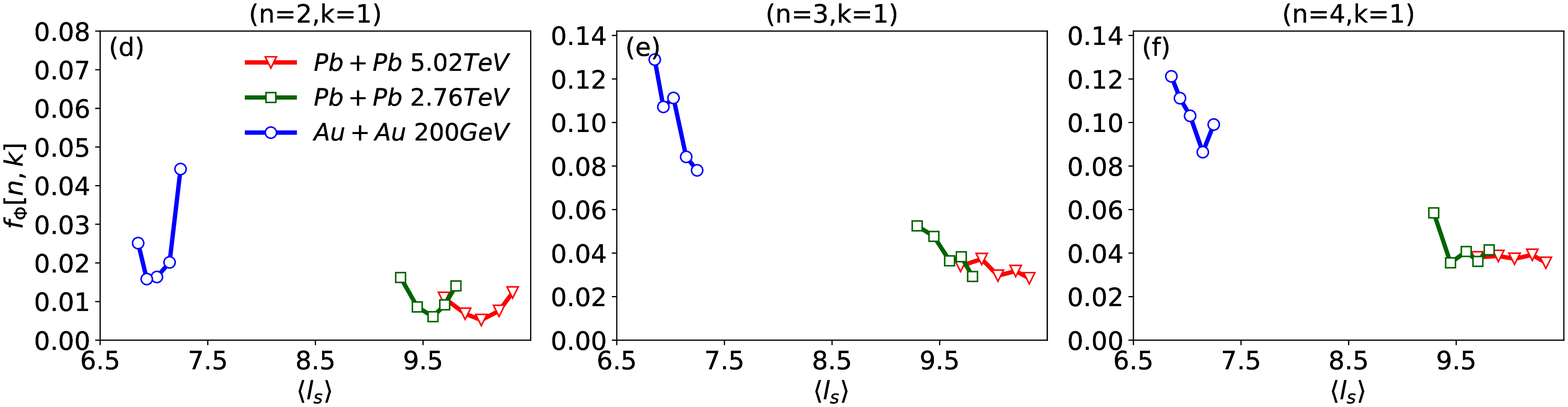}
\includegraphics[width=1.02\textwidth]{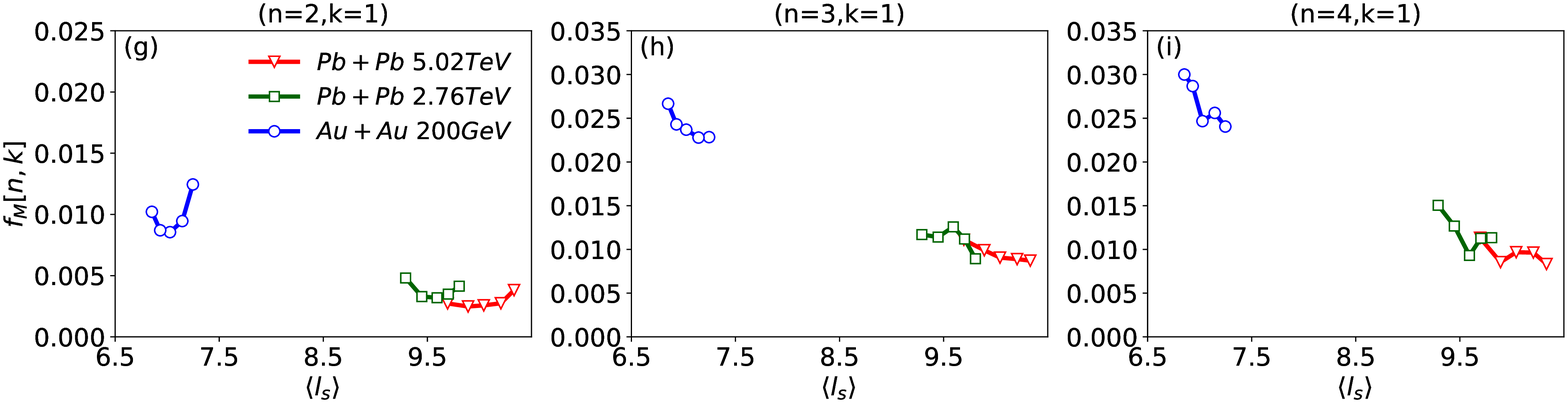}
\caption{The slope parameter $f[n,k]$, $f_M[n,k]$ and $f_\Phi[n,k]$ for $n=2,3,4$ and $k=1$ as function of the mean string length $\langle l_s \rangle$ (the collision energy and the centrality class) for Pb+Pb collisions at 5.02A TeV and 2.76A TeV at the LHC and for Au+Au collisions at 200A GeV.
In each plot (subfigure), there are three separate curves denoting three different collision energies (5.02A TeV, 2.76A TeV and 200A GeV, respectively). 
	For each curve (collision energy), there are five symbols denoting five centrality classes; from left to right, they are 40-50\%, 30-40\%, 20-30\%, 10-20\% and 5-10\%, respectively.}
\label{mean}
\end{figure*}

\begin{figure*}[thb]
\includegraphics[width=1.02\textwidth]{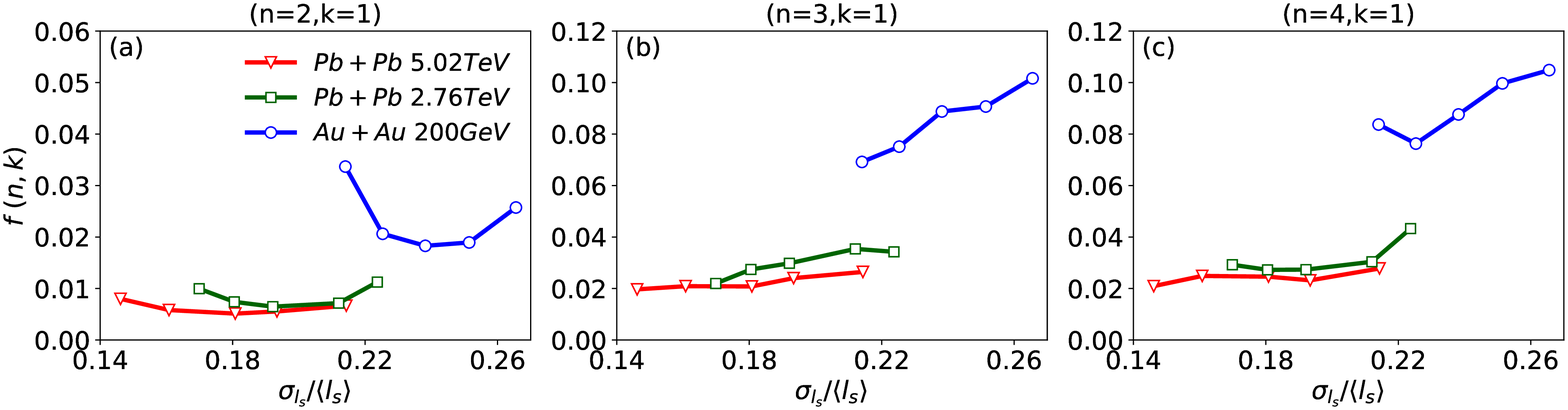}
\includegraphics[width=1.02\textwidth]{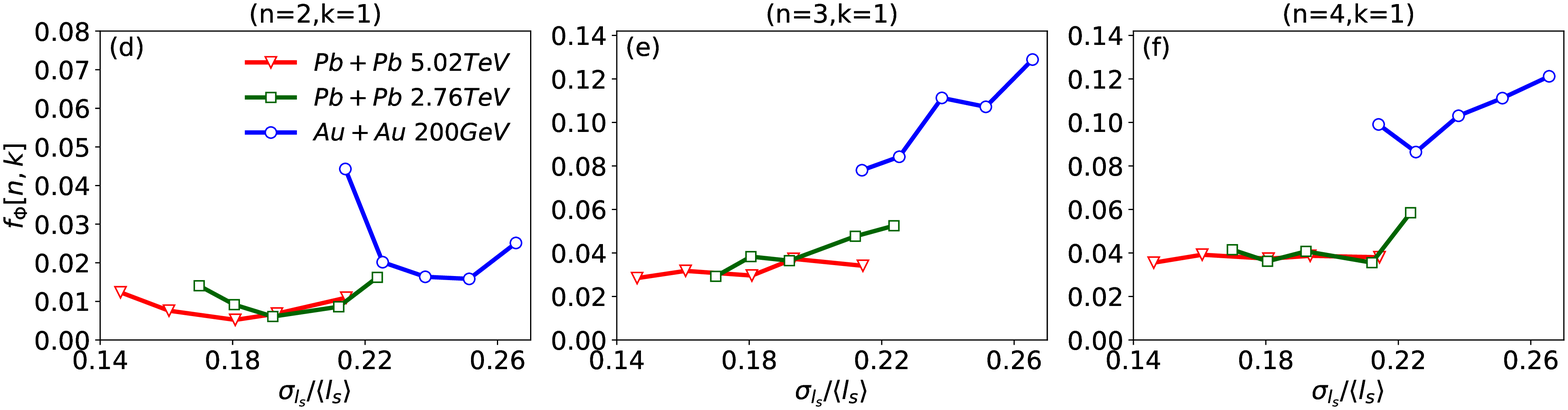}
\includegraphics[width=1.02\textwidth]{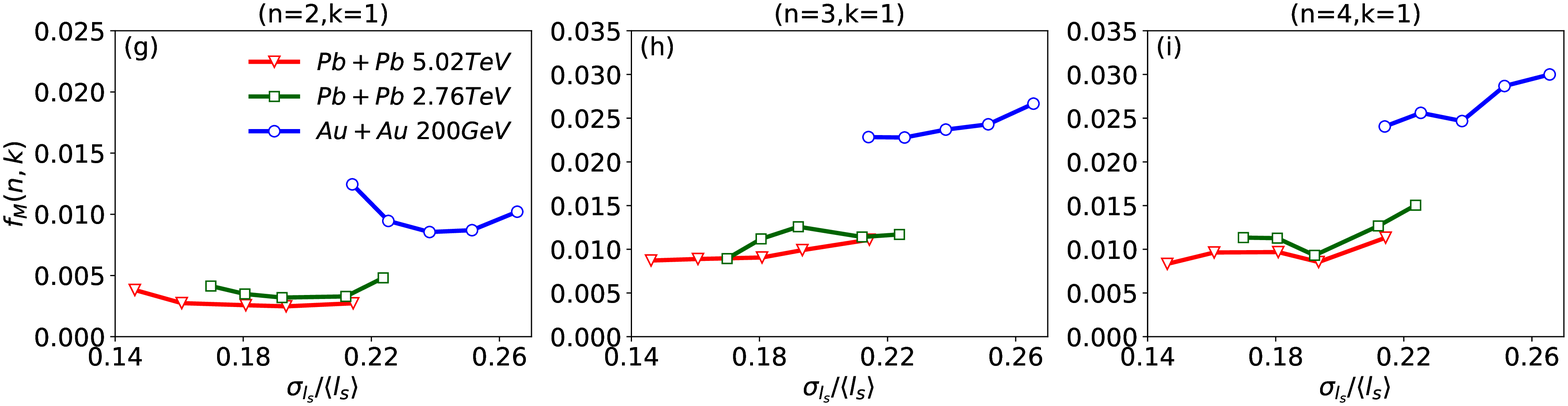}
	\caption{The slope parameter $f[n,k]$, $f_M[n,k]$ and $f_\Phi[n,k]$ for $n=2,3,4$ and $k=1$ as function of the variance-over-mean-ratio $\sigma_{l_s}/\langle l_s\rangle$ of the string length (the collision energy and the centrality class) for Pb+Pb collisions at 5.02A TeV and 2.76A TeV at the LHC and for Au+Au collisions at 200A GeV.
In each plot (subfigure), there are three separate curves denoting three different collision energies (5.02A TeV, 2.76A TeV and 200A GeV, respectively). 
	For each curve (collision energy), there are five symbols denoting five centrality classes; from right to left, they are 40-50\%, 30-40\%, 20-30\%, 10-20\% and 5-10\%, respectively.}
\label{variance}
\end{figure*}

In this subsection, we study in more detail the collision energy dependence for the longitudinal decorrelations of the anisotropic collective flows. In Figure \ref{r234enr}, we compare the slope parameters $f[n,k]$, $f_M[n,k]$ and $f_\Phi[n,k]$ with $n=2, 3, 4$ and $k=1$ as a function of collision centrality for different collision energies. The ATLAS data on $f[n,k]$ for Pb+Pb collisions are shown for comparison. We can see that the longitudinal decorrelations of anisotropic flows are much larger at RHIC than at the LHC energies. From 5.02A TeV to 2.76A TeV Pb+Pb collisions, there is around 10-20\% increase of the slope parameters for the longitudinal decorelations.

The collision energy dependence for the longitudinal decorrelations of anisotropic flows can be traced back to the fluctuations of the initial states of relativistic heavy-ion collisions.
To see this more clearly, we perform a more detailed analysis for the initial conditions generated by the AMPT model (in which the string melting mechanism is utilized to convert strings to partons).
More specifically, we use the scikit-learn package to extract the mean length of the strings in the AMPT model \cite{Pang:2016akb}.
First, we divide the initial partons into different clusters according to their transverse positions using the k-means algorithm (note that each cluster is associated with a wounded nucleon).
Then, the partons in the same cluster are arranged in longitudinal direction according to their pseudorapidities, and therefore form a string or a string-like object.
The length of each string is estimated using the difference between maximum and minimum rapidities of the initial patrons in the same cluster.
Figure \ref{length} shows the analysis results for the AMPT initial conditions for three typical central collision events in three different collision energies at the LHC and RHIC.
For the sake of easy display, we only show 20 clusters (strings) is each plot. 
One can see that when the collision energy increases, the lengths of clusters (strings) increase.
For higher collision energies, since the mean lenghts of strings are longer and the boost invariance works better, the longitudinal fluctuation effects and the decorrelations of anisotropic flows should be smaller.

In Figure \ref{mean}, we show the slope parameters $f[n,k]$, $f_M[n,k]$ and $f_\Phi[n,k]$ with $n=2, 3, 4$ and $k=1$ for different collision centralities and collision energies (both are quantified by the mean length of the strings $\langle l_s\rangle$) at the LHC and RHIC.
In each plot (subfigure), there are three curves which represent the results for three different collision energies: 5.02A TeV, 2.76A TeV and 200A GeV, respectively. 
For each curve (collision energy), there are five symbols (circles, squares or triangles) which represent the results for five centrality classes: 40-50\%, 30-40\%, 20-30\%, 10-20\% and 5-10\% (from left to right).
One can see that for a given collision energy, the lengths of the strings slightly increase from peripheral collisions to central collisions.
From the figure, we can see that for the decorrelation of the elliptic flow $v_2$, the slope parameters not only depend on the collision energy (the string lengths), but also on the collision centrality (this is because the initial elliptic collision geometry plays an important role in developing the elliptic flow $v_2$).
For $n=3, 4$, the collision geometry effect is small, and the longitudinal decorrelation of anisotropic flows mainly depend on the mean lengths of the strings.

It is also interesting to study the dependence of the longitudinal decorrelations on the fluctuations of the initial string structures (lengths). 
In Figure \ref{variance}, we show the slope parameters $f[n,k]$, $f_M[n,k]$ and $f_\Phi[n,k]$ ($n=2, 3, 4$ and $k=1$) as a function of the variance-to-mean-ratio of the string lengths $\sigma_{l_s}/\langle l_s\rangle$. 
Similar to Figure \ref{mean}, there are three curves in each subfigure (plot), representing three collision energies (5.02A TeV, 2.76A TeV and 200A GeV, respectively). 
Also there are five symbols in each curve, denoting five different centrality classes. Note that the centrality classes are 40-50\%, 30-40\%, 20-30\%, 10-20\% and 5-10\% from right to left, different from Figure \ref{mean}).
One can see that the variance-to-mean-ratio of the string lengths is typically larger for lower collision energies and more peripheral collisions. 
Again, the decorrelation of elliptic flow depends not only on the collision energy, but also on the collision centrality (the initial collision geometry), thus shows a complex dependence on the variance of the string lengths. 
As for triangular and quadranglar flows, while there are some fine structure in Figure \ref{variance}, we can still see an overall trend that the longitudinal decorrelations of $v_3$ and $v_4$ typically increase with increasing $\sigma_{l_s}/\langle l_s\rangle$ ratio (i.e., lower collision energies and more peripheral collisions).
These results indicate that the longitudinal decorrelations of anisotropic collective flows can be directly used to probe the longitudinal structures of the initial states in relativistic heavy-ion collisions.

\section{Summary}

We have performed a systematic study on the longitudinal decorrelations of anisotropic collective flows in relativistic heavy-ion collisions at the LHC and RHIC energies.
The CLVisc (ideal) (3+1)-dimensional hydrodynamics model is utilized to simulate the dynamical evolution of the QGP fireball, and the initial conditions of the hydrodynamics simulation is obtained using the AMPT model.
A detailed analysis has been performed on the longitudinal decorrelations of flow vectors, flow magnitudes and flow orientations (event planes) for elliptic, triangular and quadrangular flows.
We found that pure flow magnitudes have smaller longitudinal decorrelations than pure flow orientations, and the decorrelation of flow vectors is a combined effects of the decorrelations of flow magnitudes and orientations.
Due to initial elliptic collision geometry, the longitudinal decorrelation of elliptic flow exhibits a strong and non-monotonic centrality dependence: smallest decorrelation effect in mid-central collisions.
On the other hand, the longitudinal decorrelations of triangular and quadrangular flows show a weak dependence on the collision centrality: a slight increase for the longitudinal decorrelation from central to peripheral collisions. This is due to the fact that the systems in more peripheral collisions are smaller and thus have larger (longitudinal) fluctuation effects.

Our numerical results from event-by-event hydrodynamics simulations for Pb+Pb collisions at 5.02A TeV and 2.76A TeV at the LHC are in good agreement with the available ATLAS data.
Our predictions for Au+Au collisions at 200A GeV at RHIC show much larger longitudinal decorrelation effects as compared to Pb+Pb collisions at the LHC.
To track back the origin of the longitudinal decorrelation of anisotropic flows, we further analyze the longitudinal structures of the AMPT initial conditions.
We found that the final-state longitudinal decorrelation effects are strongly correlated with the mean lengths of the initial string structures in the AMPT model.
While the longitudinal decorrelations of elliptic flow show a non-monotonic centrality dependence due to the initial elliptic collision geometry, the decorrelation effects for triangular and quadratic flows are typically larger for lower collision energies and in more peripheral collisions due to shorter lengths of the string structures in the AMPT initial states. 
This study constitutes an important contribution to our current understanding of the initial state fluctuations, especially the fluctuations in the longitudinal directions, and their manifestations in the final states in relativistic heavy-ion collisions.

\section*{ACKNOWLEDGMENTS}

This work is supported in part by the Natural Science Foundation of China (NSFC) under grant Nos. 11775095, 11375072 and 11221504, by the Major State Basic Research Development Program in China (No. 2014CB845404), by the Director, Office of Energy Research, Office of High Energy and Nuclear Physics, Division of Nuclear Physics, of the U.S. Department of Energy under Contract Nos. DE-AC02-05CH1123, and by the US National Science Foundation within the framework of the JETSCAPE collaboration, under grant number ACI-1550228.

\bibliographystyle{h-physrev5} 
\bibliography{refs_GYQ}   

\end{document}